\documentclass{article}
\usepackage{a4wide}
\usepackage[utf8]{inputenc}
\usepackage[T1]{fontenc}

\usepackage{natbib}
\usepackage{amsmath,amssymb}

\usepackage[nameinlink]{cleveref}
\usepackage{subfig,longtable}
\usepackage{tikz}
\definecolor{blue}{rgb}{0.181,0.641,0.819}
\definecolor{green}{rgb}{0.297,0.673,0.407}
\definecolor{orange}{rgb}{1,0.64,0.159}

\begin{document}

\title{Penalty shootouts are tough,\\but the alternating order is fair}
\author
{
 Silvan Vollmer\\ETH Zürich
 \and
 David Schoch\\GESIS
 \and
 Ulrik Brandes\\ETH Zürich
}
\date{\today}

\maketitle
\begin{abstract}
We compare conversion rates of association football (soccer) penalties
during regulation or extra time with those during shootouts.
Our data consists of roughly 50,000 penalties from
the eleven~most recent seasons in European men's football competitions.
About one third of the penalties are
from more than 1,500 penalty shootouts.  
We find that shootout conversion rates are significantly lower,
and attribute this to worse performance of shooters
rather than better performance of goalkeepers.
We also find that, statistically,
there is no advantage for either team
in the usual alternating shooting order.
These main findings are complemented by
a number of more detailed analyses.
\end{abstract}

\section{Introduction}\label{sec:intro} 

Football is a low-scoring game,
with typically less than three goals per match~\citep{gurkan:2017}. 
On average, one in three matches sees a penalty awarded,
and more than three in four of these lead to a goal.
Moreover, since their adoption
by the International Football Association Board (IFAB) in 1970,
penalty shootouts are the ultimate method of choice to break ties
in competitions with elimination rounds.
Penalties are therefore an important element of the game,
both during regulation and in shootouts.

Research into factors influencing the outcome of a penalty abounds.
It includes findings on
player nationality~\citep{brinkschulte:2020}, 
player status~\citep{jordet:2009},
penalty importance~\citep{jordet:2007},
the target zone of a shot~\citep{almeida:2016,jamil:2020,horn:2021},
run-up length, strike type, footedness, match status~\citep{jamil:2020}
and other circumstantial factors~\citep{arrondel:2019,jamil:2020,horn:2021}.
Specifically for penalty shootouts,
the impact of a team's league~\citep{krumer:2020},
team strength and match venue~\citep{wunderlich:2020}
have been considered.

Our focus, however,
is not on strategies and factors influencing success,
but on empirical regularities in penalty conversion rates
under in-game and shootout conditions.
In other words, we do not intend to identify success factors,
but to assess whether there are outcome differences
associated with the two major settings in which penalties are taken.

A majority of prior studies indeed observe a higher conversion rate
for in-game penalties~\citep{brinkschulte:2020,brinkschulte:2023,mcgarry:2000}
compared to shootout penalties~\citep{dalton:2015}.
Based on a much larger and more recent data set,
we not only confirm this finding,
but reveal that this is almost entirely due to a larger number of misses.

A hotly debated issue for shootouts is
whether the alternating shooting order favors the team going first.
A number of models suggest that, in theory, the current order is unfair,
because a team lagging behind faces higher pressure
so that the beginning team enjoys a first-mover advantage
\citep{csato:2020,lambers:2020,vandebroek:2016,csato:2022,brams:2018}.
Empirically, however, the evidence is more controversial.
While several studies find a bias
\citep{apesteguia:2010,palacios:2014,rudi:2020,dasilva:2018}
others do not~\citep{kocher:2012,kassis:2021,arrondel:2019,santos:2022}.
The most recent study based on all shootouts in all major tournaments
of the past fifty years sides with the latter group~\citep{santos:2022}. 
From even more, and crucially also more recent data 
we conclude that, at least in men's football,
no discernible bias results from the alternating order.

The remainder of this paper is organized as follows.
In \Cref{sec:data}, we introduce the data on which our study is based.
Our analyses in \Cref{sec:results} are focused on
differences between in-game and shooutout penalty outcomes, 
but include separate analyses of both conditions
and of players subjected to both.
We offer conclusions in \Cref{sec:conclusion}.

\section{Data}\label{sec:data}

We obtained data on penalties 
in 115~major national and international football competitions throughout Europe.
The data has been scraped from Transfermarkt, a commercial website
with comprehensive and up-to-date information on players, clubs, and competitions,
in July~2023.\footnote{Note to reviewers:
We intend to make the cleaned and formatted dataset available
in an open repository for replication and further study.}

Only penalties awarded in one of the past eleven~seasons
(2012-13 to 2022-23, starting with EURO~2012)
and in men's competitions are considered.
Both these restrictions serve to increase coherence and avoid systematic biases,
as it has been suggested, for instance, that the mechanisms governing shootout results
may differ substantially in women's football~\citep{santos:2022}.
Included are competitions run by either UEFA
or one of the 55~football associations that are members of UEFA.
A total of 54~leagues constitute the highest level of professional football
in each of the associations, if any,\footnote{There is no league in Liechtenstein.}
and a total of 54~national cup competitions add some breadth.%
\footnote{Transfermarkt does not cover
cup competitions for North Macedonia and for Montenegro,
but for England we include two, the FA~Cup and League Cup.}
In addition to six international competitions organized by UEFA
(European Championship, Nations League, Champions League,
Europa League, Conference League, and Super Cup),
we include the FIFA World Cup Finals of~2014, 2018, and~2022,
where many of the non-European teams field players who play for European clubs.
With almost every league accompanied by a cup,
more than half of the competitions feature elimination rounds
in which ties are ultimately resolved by penalty shootouts.

Scraping match summaries
from the 115~competitions above
over the past eleven years
yielded 1,114~competition-season pairs with a total of 67,310~penalties.
We refer to this raw data as dataset~D0.
The data published by Transfermarkt is not necessarily from official records,
and we cannot rely on its accuracy, completeness, or consistency. 
By removing data of dubious quality from~D0, 
we attempt to mitigate possible effects of apparent or suspected reporting issues.

\subsection{Penalty occurrence rates (D1)}

In a first step, the raw dataset D0 is filtered
for apparent inconsistencies and suspected reporting biases.

While in-game (i.e., regulation or extra time) penalties
are subject to a wider variety of inconsistencies,
the rules for shootout penalties make it easier to spot such inconsistencies.
The following two sets of filters have been applied
to remove potentially problematic cases:
\begin{description}
\item[In-game penalties.]\quad
By querying in-game penalties for the same team within minutes,
we identified 105~in-game penalties that were deemed double entries,
generally because a goal was scored from a rebound.
After manually consolidating these,
we turned our attention to potential under-reporting.
In niche competitions especially,
penalties are sometimes included only if they led to a goal,
or goals are recorded without indication that they resulted from a penalty.
We therefore applied frequency tests to eliminate seasons of competitions
that featured significantly fewer penalties than the others.
Based on the respective numbers of matches and penalties,
we estimated occurrence rates of
0.2835~penalties during regulation
and 0.0763~during extra time. 
A one-sided Poisson test at the 1\%~level
flagged 108~seasons across 45~different competitions 
in which in-game penalties may have been under-reported.
We therefore dropped these seasons from the dataset, 
resulting in 2,794~fewer in-game penalties.

\item[Shootouts.]\quad
Shootouts are subject to additional regulations.
According to the Laws of the Game~\citep{laws:2022}
five penalties each are taken in alternating order between the two teams. 
A shootout is terminated
if a team can no longer level the score with the remaining penalties,
and it is continued beyond the first ten attempts
with one additional penalty for each team
as long as the score remains tied.
An entire shootout is therefore removed from the dataset,
if the order between the teams is not alternating,
the number of penalties is less than six or an odd number larger than ten,
all of its penalties resulted in a goal,
or the final score is a draw. 

A combined total of 4,124~penalties were removed
because they occurred in a shootout with inconsistent data.
\end{description}

Having removed shootouts and seasons of competitions 
with clearly or potentially inconsistent reporting,
we are left with a initial dataset, D1, of 60,287~penalties.
From this, we proceed with additional quality checks
to extract the two main dataset used in our study. 
They differ by the level of detail required for the outcome of a penalty.

\subsection{Two-way outcomes:\enskip goal or no goal (D2)}

For analyses that require only knowledge about whether a penalty kick
resulted in a goal or not,
we still need to ensure that unsuccessful kicks are indeed reported.
In parallel to the overall rate by which penalties are awarded, 
we therefore applied an additional frequency test for in-game penalty outcomes.

Starting from the full scraped dataset~D0,
687~penalties with missing or indeterminate outcomes are excluded.
From the remaining in-game penalties,
we obtained an initial estimate of 82.06\% for the in-game penalty conversion rate
and applied a binomial test to filter out time periods
with seemingly lower reporting standards.
A total of 7,637~penalties were taken in seasons
with significantly different conversion rates (5\%~level).
After excluding these, 
the initial estimate for the in-game conversion rate is revised to~82.00\%.
A second, stricter, binomial test is then applied
to flag systematic reporting issues for entire competitions
in which the accumulated conversion rates over all eleven~seasons
deviate significantly on the 1\%~level.
Dropping these competitions excludes another 1,604~penalties.
The two tests combined thus led to the exclusion of 9,241~penalties
from 231~seasons across 64~different competitions.

Taking into account that some of the flagged penalties have already been marked
by the Poisson test for in-game penalty occurrence (rather than conversion) rates,
the resulting dataset~D2 to be used in analyses involving binary outcomes
contains a total of 51,007~penalties.

\subsection{Three-way outcomes:\enskip goals, saves, and misses (D3)}

In some analyses, we are also interested in the different rates
by which penalties are saved by the goalkeeper or missed by the kicker.
The differentiation between these two non-goal outcomes
presents additional sources of error in the scraped data.

Starting from the scraped data~D0, we again eliminated 
the 687~penalties with incomplete outcomes
and an additional 37~non-goal penalties
without distinction between the two possible reasons.
The outcome labels for the remaining 62,357~penalties
are normalized to either goal, saved, or missed
with estimated rates of 
82.13\% (goal), 12.96\% (saved), and 4.91\% (missed) for in-game penalties,
and 74.64\% (goal), 14.30\% (saved), and 11.06\% (missed) for shootouts.
Multinomial tests are then applied to identify further reporting biases
such as non-goal penalties invariably reported as missed.

The season-wise 5\%-level significance test flagged a combined number of
7,976~in-game penalties and 3,793~shootout penalties.
Revised outcome-rate estimates after their exclusion are
82.55\% (goal), 12.85\% (saved), and 4.60\% (missed) for in-game penalties,
and 74.92\% (goal), 14.97\% (saved), and 10.11\% (missed) for shootouts.
The subsequent 1\%-level significance test
for competitions over the entire eleven~seasons
flagged a combined number of 2,058~in-game penalties and 921~shootout penalties.

In total, 14,748~penalties from 279~seasons across 82~different competitions
were flagged.
Excluding these penalties from the initial dataset,
again in combination with the Poisson test for penalty occurrence rates,
yields a dataset~D3 comprised of 45,402~penalties
with plausibly recorded three-way outcomes.

\begin{table}
\caption{Description of penalty datasets.
Data scraped from Transfermarkt for the 2012/13--2022/23 seasons
has been filtered to exclude seasons and competitions
with implausible occurrence rates and outcome distributions
attributed to inconsistent reporting.}
\label{tab:datasets}
\begin{tabular}{l|r|rrrr}
dataset & penalties & regulation & extra time & in-game & shootout\\\hline
D0:\enskip as scraped		& 67,310 & 43,846 & 445 & 44,291 & 23,019\\	
D1:\enskip penalty occurrences	& 60,287 & 40,961 & 423 & 41,384 & 18,903\\\hline 	
D2:\enskip two-way outcomes	& 51,007 & 31,781 & 323 & 32,105 & 18,903\\ 	
D3:\enskip three-way outcomes	& 45,402 & 30,908 & 305 & 31,213 & 14,189\\ 	
\end{tabular}
\end{table}

After cleaning and plausibility filtering 
we obtain the two main datasets, D2 and~D3 (see \Cref{tab:datasets})
with the following attributes included for each penalty:
\begin{itemize}
\item competition and season 
\item name and result of the corresponding match
\item in-game (regulation or extra time), or shootout
\item time (minute for in-game, position in shooting order for shootout)
\item penalty taker, goalkeeper, and their respective clubs
\item score at which the penalty was taken
\item outcome: goal/no goal in D2 and goal/save/miss in D3.
\end{itemize}
In both~D2 and~D3,
roughly one third of the penalties are from shootouts.
Home teams are awarded $55.82\%$~($55.96\%$)
of all in-game penalties in dataset~D2~(D3),
for a strikingly high advantage of more than eleven~percentage points.
Across all 115~competitions and eleven~seasons
we identified 21,948~(18,815)~unique penalty takers
and 7,194~(6,808)~unique goalkeepers in the two-way (three-way) dataset.%
\footnote{For~1,117~(881)~penalties in~D2~(D3), information on the goalkeeper is missing.}

\section{Analysis}\label{sec:results}

We start this section with some general statistics on penalty outcomes, 
and compare conversion rates in-game and during shootouts in particular.
Since we find that shooter performance is significantly lower during shootouts,
further analysis is directed at
player selection (shootout penalties have to be taken by different players),
fatigue (shootouts take place at the end of a match),
and shooting order (increased pressure on the team going second in a shootout)
to investigate possible reasons for this observation.

\subsection{In-game vs.\ shootout}\label{sec:outcomes}

Previous research reports penalty conversion rates between 68\% and 85\%
\citep{almeida:2016,brinkschulte:2020,mcgarry:2000,dalton:2015,jordet:2007,jamil:2020,jordet:2009,apesteguia:2010,palacios:2014,horn:2021,bar-eli:2009,lopez:2007,white:2013,dohmen:2008,furley:2016,brinkschulte:2023}
and most of the time finds higher conversion rates for in-game penalties
\citep{brinkschulte:2020,brinkschulte:2023,mcgarry:2000}.
The opposite is found in one study~\citep{dalton:2015},
albeit based on data from only six tournaments.
None of the data sets in any of these studies included more than 10,000 penalties,
and some of them span multiple decades.

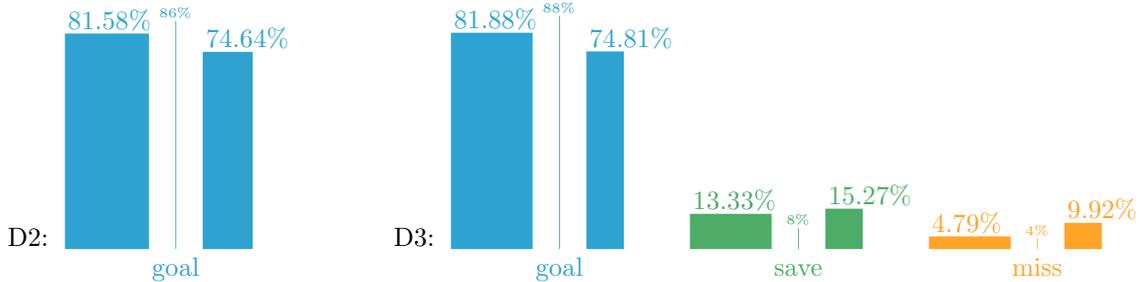
\begin{figure} 
\centerline{\begin{tikzpicture}[x=1pt,y=1pt,inner sep=1pt]
\fill[blue]
  +(-5,0) node[black,above left,anchor=south east]{D2:}
  +(0,81.58) node[above,anchor=south west]{$81.58\%$} rectangle ++(31.781,0)
  ++(10,0) +(0,-10) node[anchor=base]{goal}
  +(0,86.38) node[above,anchor=south]{\tiny$86\%$} rectangle ++(0.323,0)
  ++(10,0)
  +(0,74.64) node[above,anchor=south west]{$74.64\%$} rectangle ++(18.903,0)
  coordinate (C);

\fill[blue] (C) ++(75,0)
  +(-5,0) node[black,above left,anchor=south east]{D3:}
  +(0,81.88) node[above,anchor=south west]{$81.88\%$} rectangle ++(30.908,0)
  ++(10,0) +(0,-10) node[anchor=base]{goal}
  +(0,88.20) node[above,anchor=south]{\tiny$88\%$} rectangle ++(0.305,0)
  ++(10,0)
  +(0,74.81) node[above,anchor=south west]{$74.81\%$} rectangle ++(14.189,0)
  coordinate (C);

\fill[green] (C) ++(25,0)
  +(0,13.33) node[above,anchor=south west]{$13.33\%$} rectangle ++(30.908,0)
  ++(10,0) +(0,-10) node[anchor=base]{save}
  +(0,7.87) node[above,anchor=south]{\tiny$8\%$} rectangle ++(0.305,0)
  ++(10,0)
  +(0,15.27) node[above,anchor=south west]{$15.27\%$} rectangle ++(14.189,0)
  coordinate (C);

\fill[orange] (C) ++(25,0)
  +(0,4.79) node[above,anchor=south west]{$4.79\%$} rectangle ++(30.908,0)
  ++(10,0) +(0,-10) node[anchor=base]{miss}
  +(0,3.93) node[above,anchor=south]{\tiny$4\%$} rectangle ++(0.305,0)
  ++(10,0)
  +(0,9.92) node[above,anchor=south west]{$9.92\%$} rectangle ++(14.189,0);
\end{tikzpicture}}
\caption{Relative frequency of outcomes in two-way and three-way data sets~D2 and~D3
split by match period (regulation, extra time, shootout).
Widths of bars are proportional to percentage of penalties in the respective period,
showing that numbers during extra time are too small to be relevant.
We therefore combine regulation and extra time into in-game penalties.}
\label{fig:outcome_all}
\end{figure}

From our larger and more recent data set,
we conclude that in-game penalties
are converted significantly more often than shootout penalties,
with a difference of seven percentage points. 
Roughly speaking,
four out of five penalties are converted successfully during a match,
but only three out of four penalties during shootouts.
See \Cref{fig:outcome_all} for details.

Interestingly, the 7-point drop in shooting performance
is accounted for by a 5-point increase in misses,
and only a 2-point increase in saves.
The immediate effect of goalkeeper performance,
often considered to be the decisive element in a shootout,
is thus more limited than it may appear to be.
We add another layer of interpretation to this in \Cref{sec:selection}
and are not aware of any prior research noting this.

\begin{table}
\caption{Outcome rates for leagues in D3.
Rates do not differ between the Big-5 and other leagues.
All penalties have been taken during regulation 
as there are neither extra time nor shootouts in these competitions.}
\centerline{\begin{tabular}{l|rrr|rrrrr}
 & \multicolumn{3}{c|}{D2} & \multicolumn{5}{c}{D3} \\ 
 & leagues & penalties & goal & leagues & penalties & goal & save & miss\\\hline
  Big-5 	&   5 &  4,500 & 81.04\% &  5 &  4,034 & 81.26\% & 13.78\% & 4.96\%\\
  other leagues &  48 & 19,820 & 80.95\% & 47 & 19,485 & 81.14\% & 13.92\% & 4.93\%\\\hline
  combined	&  53 & 24,320 & 80.97\% & 52 & 23,519 & 81.16\% & 13.90\% & 4.94\%\\
\end{tabular}}
\label{league outcomes}
\end{table} 

A comparison across leagues shows that outcome rates
do not depend on the level at which leagues are operating.
The first tiers in England, Spain, Italy, Germany, and France
are generally considered to be the top leagues within UEFA.
This is corroborated by the averages of the UEFA Country Coefficients
from 2012/13 to 2022/23.
As can be seen in \Cref{league outcomes},
outcome rates are essentially the same in these and all other leagues.
It appears that, on average,
the quality of penalty takers and goalkeepers vary jointly.

\begin{table}
\caption{Outcome rates for cup competitions in D3.
Rates fluctuate where numbers are small,
but overall in-game goal probabilities are even higher than in league competitions.}
\centerline{\begin{tabular}{lr|rrrr|rrrr}
&&
\multicolumn{4}{c|}{in-game (regulation and extra time)} &
\multicolumn{4}{c}{shootout}\\
 & cups &
penalties & goal & save & miss &
penalties & goal & save & miss \\\hline
national & 46 &
 6,615 & 84.93\% & 11.07\% & 4.01\% &
13,739 & 74.79\% & 15.21\% & 9.99\% \\
international (clubs) & 4 &
   863 & 80.88\% & 13.21\% & 5.91\% &
   256 & 79.30\% & 15.23\% & 5.47\% \\
international (selections) & 3 & 
   216 & 80.09\% & 13.43\% & 6.48\% &
   194 & 70.10\% & 19.59\% &10.31\% \\\hline
combined & 53 & 
 7,694 & 84.34\% & 11.37\% & 4.29\% &
14,189 & 74.81\% & 15.27\% & 9.92\% \\
\end{tabular}}
\label{cup outcomes}
\end{table} 

In national cup competitions,
participating teams are from different league tiers within a country, 
so that quality levels of penalty takers and goalkeepers are more varied.
The outcome rates shown in \Cref{cup outcomes} suggest
that this results in an advantage for penalty takers, 
because in-game goal rates are higher than in leagues.
While we suspect that this is because better teams with better players 
have more possession and see more penalties awarded to them, 
we did not investigate this further.%
\footnote{A more detailed analysis on the dyadic level (kicker vs.\ goalkeeper)
is planned for future work.}
Instead, we provide a few rankings of individuals and teams in the next subsection.

The important observation is that conversion rates are lower in shootouts
even when comparing to in-game penalties restricted to cup competitions.
If anything, the effect is more pronounced.

\subsection{Kickers, Goalkeepers, Teams}

Since we collected outcome data only,
we make no attempt at explaining patterns of success.
It is nevertheless interesting to see how top performers compare
to the baselines established in the previous section.
In addition, these qualitative evaluations provide the context
that facilitates interpretation of our subsequent analyses.

We have therefore compiled lists of penalty takers, goalkeepers, and teams
that appear most frequently in our data.
Filtering thresholds for the number of penalties are set such that
at least 50~players or teams are selected into a list. 
In case of a tie for the 50th rank,
all players or teams with the same number of appearances are included.

\subsubsection{Penalty takers}

\begin{figure}
\edef\penaltytakers{%
Sölvi Vatnhamar/30/96.6666666666667,
Marcin Robak/33/93.9393939393939,
Besnik Krasniqi/32/93.75,
Bibras Natcho/46/93.4782608695652,
Harry Kane/49/91.8367346938776,
Romelu Lukaku/35/91.4285714285714,
Radamel Falcao/32/90.625,
Bruno Fernandes/50/90,
Dusan Tadic/40/90,
Bas Dost/30/90,
Lior Refaelov/30/90,
Neymar/30/90,
Filip Starzynski/46/89.1304347826087,
Andrej Kramaric/43/88.3720930232558,
Zakaria Beglarishvili/34/88.2352941176471,
Domenico Berardi/41/87.8048780487805,
Jóannes Bjartalid/41/87.8048780487805,
Karim Benzema/31/87.0967741935484,
Mario Balotelli/38/86.8421052631579,
Jorginho/53/86.7924528301887,
Mikel Oyarzabal/30/86.6666666666667,
Thomas Müller/30/86.6666666666667,
Wissam Ben Yedder/44/86.3636363636364,
Robert Lewandowski/51/86.2745098039216,
Ivan Krstanovic/50/86,
Clayton Failla/35/85.7142857142857,
Alexandre Lacazette/41/85.3658536585366,
Guillaume Hoarau/34/85.2941176470588,
João Pedro/34/85.2941176470588,
Flávio Paixão/40/85,
Michael Liendl/39/84.6153846153846,
Iago Aspas/32/84.375,
Andrea Belotti/31/83.8709677419355,
Lorenzo Insigne/37/83.7837837837838,
Cristiano Ronaldo/73/83.5616438356164,
Bruno Petkovic/30/83.3333333333333,
Eden Hazard/30/83.3333333333333,
Max Gradel/30/83.3333333333333,
Jamie Vardy/33/81.8181818181818,
Edinson Cavani/41/80.4878048780488,
Teun Koopmeiners/35/80,
Patrick Hoban/32/78.125,
Lionel Messi/44/77.2727272727273,
James Tavernier/52/76.9230769230769,
Ciro Immobile/56/76.7857142857143,
Sergio Agüero/37/75.6756756756757,
Paulinho/36/75,
Memphis Depay/31/74.1935483870968,
Mijo Caktas/38/73.6842105263158,
Klaemint Olsen/30/73.3333333333333,
Dani Parejo/30/66.6666666666667}
\centerline{%
\subfloat[goal percentages of frequent kickers]
 {\begin{tikzpicture}[scale=0.15,outer sep=3pt]
  \draw[gray,dashed] (28,60) node[left]{$60\%$} coordinate (B) -- node[above]{number of penalties} (80,60);
  \draw[gray,dashed] (30,60) -- node[blue,yshift=5,near start,above,sloped]{goal percentage} (30,100);
  \foreach \x in {30,50,73} {\draw[gray] (\x,60) -- +(0,-2) node[below]{\x};}
  \draw[blue] (28,90) node[left]{$90\%$} -- (80,90) node[above left]{\tiny Top10};
  \draw[blue,dashed]
    (28,85) node[left]{$85\%$} -- (80,85)
    (28,79) node[left]{$79\%$} -- (80,79);
  \foreach \n/\p/\r in \penaltytakers { \fill (\p,\r) circle (0.75); } 
  \draw (30,96.67) node[above right]{Vatnhamar};
  \draw (49,91.84) node[above right,yshift=-3]{Kane};
  \draw (46,93.48) node[above]{Natcho};
  \draw (50,90.00) node[right]{Fernandes};
  \draw (73,83.56) node[below,xshift=-10]{\scriptsize Cristiano Ronaldo};
  \draw (56,76.79) node[right]{\scriptsize Ciro Immobile};
  \draw (44,77.72) node[below right,xshift=-5,yshift=-2]{\scriptsize Lionel}
                   node[below right,xshift=-5,yshift=-10]{\scriptsize Messi};
  \draw (30,66.67) node[right]{\scriptsize Dani Parejo}; 
  \end{tikzpicture}
  \label{top50 players}
}
\qquad
\subfloat[Top10+2 in goal percentage]
{\begin{tabular}[b]{lrr}
kicker & \llap{penalties}  & goal \\ \hline
S{\o}lvi Vatnhamar & 30 & 96.67\%\\
Marcin Robak & 33 & 93.94\%\\
Besnik Krasniqi & 32 & 93.75\%\\
Bibars Natcho & 46 & 93.48\%\\
Harry Kane & 49 & 91.84\%\\
Romelu Lukaku & 35 & 91.43\%\\
Radamel Falcao & 32 & 90.63\%\\
Bruno Fernandes & 50 & 90.00\%\\
Du{\v s}an Tadi\'c & 40 & 90.00\%\\
Bas Dost & 30 & 90.00\%\\
Neymar & 30 & 90.00\%\\
Lior Refaelov & 30 & 90.00\%\\
\end{tabular}
\label{top10 players}
}}
\caption{Most frequent penalty kickers in dataset~D2.
There are 51~players with 30~or more attempts.
Across all competitions and match periods,
$79\%$ of penalties in~D2 result in a goal.}
\label{top players}
\end{figure}
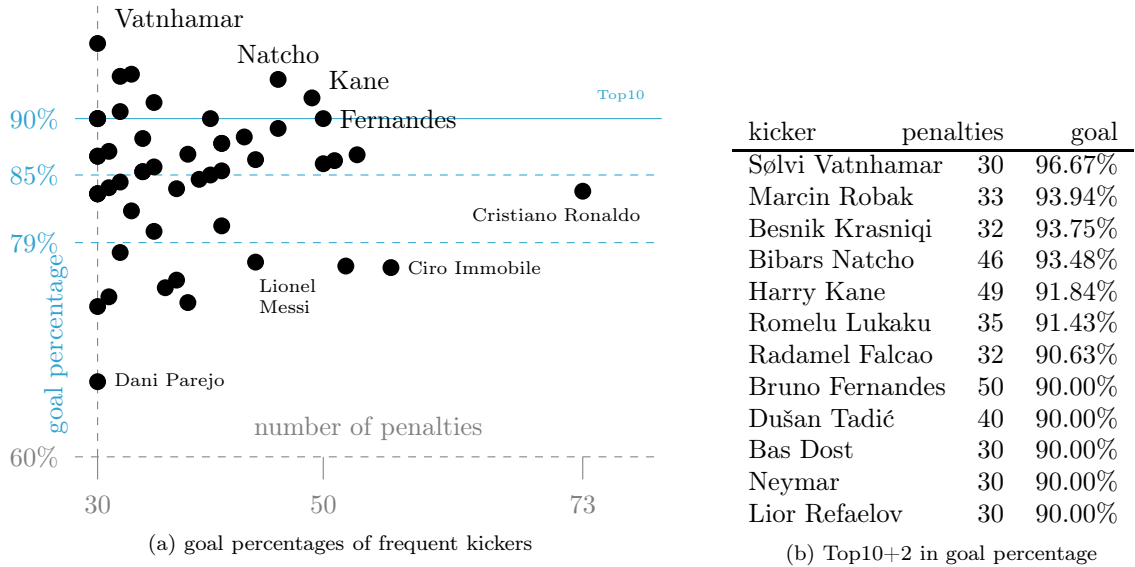

It would seem rational that the best kickers appear most often at the mark,
because better teams win more penalties and have better kickers execute them.
A threshold of at least 30~penalties yields 51~penalty takers
appearing most frequently in dataset~D2.
We compare their success rates in \Cref{top50 players}.
Their collective conversion rate of 84.77\%
compares favorably to the 78.81\% of all others
(for an overall total of 79.04\%),
suggesting that teams do tend to select from their most dependable kickers.

Over the past eleven~seasons,
S{\o}lvi Vatnhamar is the most successful high-volume penalty taker,
converting all but one of his 30~penalties
for V{\'\i}kingur G{\o}ta in the Faroe Islands Premier League.
A different player who converted all but one of his attempts,
but does not appear in the above list, is Max Kruse.
For five different clubs he has taken a total of 27~penalties 
and therefore did not make the threshold.

The player who took by far the most penalties, Cristiano Ronaldo,
has a success rate just below the average of all frequent kickers,
whereas another frequent penalty taker, Lionel Messi,
succeeded just below the overall average. 
For a player taking many penalties, 
the goal percentage of Dani Parejo is astonishingly low (20~out of~30).

Above we do not distinguish between in-game and shootout penalties, 
because players have too few shootout appearances for analysis.
The maximum is attained by Jorginho,
who has taken eleven~shootout penalties (and converted nine),
and for the next three players
(James Maddison, Mason Mount, and Paulinho)
this number is already down to eight.

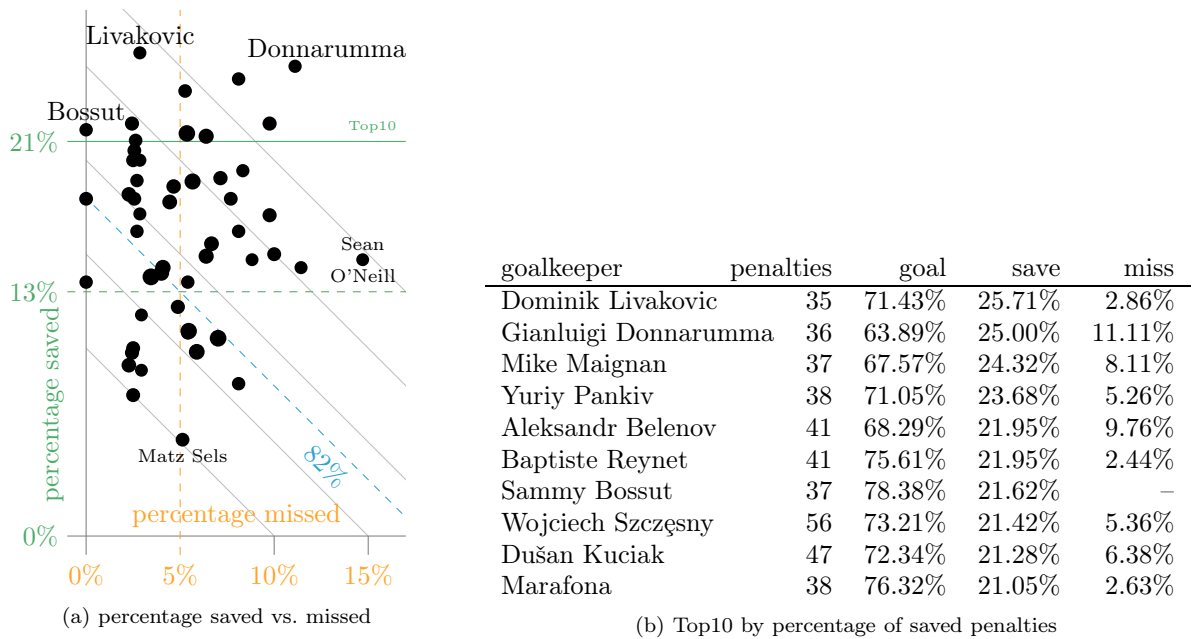
\begin{figure}[t!]
\edef\goaliesingame{%
Dominik Livakovic/35/71.4285714285714/25.7142857142857/2.85714285714286,
Gianluigi Donnarumma/36/63.8888888888889/25/11.1111111111111,
Mike Maignan/37/67.5675675675676/24.3243243243243/8.10810810810811,
Yuriy Pankiv/38/71.0526315789474/23.6842105263158/5.26315789473684,
Aleksandr Belenov/41/68.2926829268293/21.9512195121951/9.75609756097561,
Baptiste Reynet/41/75.609756097561/21.9512195121951/2.4390243902439,
Sammy Bossut/37/78.3783783783784/21.6216216216216/0,
Wojciech Szczesny/56/73.2142857142857/21.4285714285714/5.35714285714286,
Dusan Kuciak/47/72.3404255319149/21.2765957446809/6.38297872340426,
Marafona/38/76.3157894736842/21.0526315789474/2.63157894736842,
Jeroen Zoet/39/76.9230769230769/20.5128205128205/2.56410256410256,
Heinz Lindner/35/77.1428571428572/20/2.85714285714286,
Simon Mignolet/40/77.5/20/2.5,
Lukasz Fabianski/36/72.2222222222222/19.4444444444444/8.33333333333333,
Oliver Baumann/42/73.8095238095238/19.047619047619/7.14285714285714,
Emil Audero/37/78.3783783783784/18.9189189189189/2.7027027027027,
Romain Ruffier/53/75.4716981132076/18.8679245283019/5.66037735849057,
Benoît Costil/43/76.7441860465116/18.6046511627907/4.65116279069767,
Andrea Consigli/44/79.5454545454545/18.1818181818182/2.27272727272727,
Bartlomiej Dragowski/39/79.4871794871795/17.9487179487179/2.56410256410256,
Sergio Padt/39/74.3589743589744/17.9487179487179/7.69230769230769,
Sinan Bolat/39/82.051282051282/17.9487179487179/0,
Fernando Muslera/45/77.7777777777778/17.7777777777778/4.44444444444444,
Kevin Trapp/35/80/17.1428571428571/2.85714285714286,
Yann Sommer/41/73.1707317073171/17.0731707317073/9.75609756097561,
Manuel Neuer/37/75.6756756756757/16.2162162162162/8.10810810810811,
Pepe Reina/37/81.0810810810811/16.2162162162162/2.7027027027027,
Lukasz Skorupski/45/77.7777777777778/15.5555555555556/6.66666666666667,
Silviu Lung Jr./40/75/15/10,
Samir Handanovic/47/78.7234042553192/14.8936170212766/6.38297872340426,
Mladen Matkovic/34/76.4705882352941/14.7058823529412/8.82352941176471,
Sean O'neill/34/70.5882352941177/14.7058823529412/14.7058823529412,
Frantisek Plach/35/74.2857142857143/14.2857142857143/11.4285714285714,
Kasper Schmeichel/49/81.6326530612245/14.2857142857143/4.08163265306122,
Rui Patrício/50/82/14/4,
Alban Lafont/58/82.7586206896552/13.7931034482759/3.44827586206897,
Emiliano Viviano/37/81.0810810810811/13.5135135135135/5.40540540540541,
Ruud Boffin/37/86.4864864864865/13.5135135135135/0,
Guillaume Faivre/41/82.9268292682927/12.1951219512195/4.8780487804878,
Arkadiusz Malarz/34/85.2941176470588/11.7647058823529/2.94117647058824,
Anthony Lopes/55/83.6363636363636/10.9090909090909/5.45454545454545,
Benjamin Lecomte/37/83.7837837837838/10.8108108108108/5.40540540540541,
Marco Bizot/57/82.4561403508772/10.5263157894737/7.01754385964912,
Bernd Leno/40/87.5/10/2.5,
Hugo Lloris/51/84.3137254901961/9.80392156862745/5.88235294117647,
Juan Musso/41/87.8048780487805/9.75609756097561/2.4390243902439,
Paul Bernardoni/44/88.6363636363636/9.09090909090909/2.27272727272727,
Jonathan Joubert/34/88.2352941176471/8.82352941176471/2.94117647058824,
Erwin Mulder/37/83.7837837837838/8.10810810810811/8.10810810810811,
Steve Mandanda/40/90/7.5/2.5,
Matz Sels/39/89.7435897435898/5.12820512820513/5.12820512820513}
\centerline{\subfloat[percentage saved vs.\ missed]
 {\begin{tikzpicture}[scale=0.25]
  \begin{scope}
    \clip (0,0) rectangle (17,28);
    \foreach \x in {10,15,20,25,30} {\draw[gray!50] (\x,0) -- (0,\x);}
    \draw[blue,dashed] (18,0) -- node[below,near start,sloped]{$82\%$} (0,18);
  \end{scope}
  \draw[gray] (-1,0) node[green,left]{$0\%$} coordinate (B) -- node[orange,above]{percentage missed} (17,0);
  \draw[gray] (0,0) -- node[green,above,sloped,near start,yshift=5]{percentage saved} (0,28);
  \foreach \x in {0,10,15} {\draw[gray] (\x,0) -- +(0,-1) node[orange,below]{$\x\%$};}
  \draw[orange,dashed] (5,-1) node[orange,below]{$5\%$} -- +(0,29);
  \draw[green,dashed] (-1,13) node[left]{$13\%$} -- +(18,0);
  \draw[green] (-1,21) node[left]{$21\%$} -- +(18,0) node[above left]{\tiny Top10};
  \foreach \n/\p/\r/\s/\m in \goaliesingame { \fill (\m,\s) circle (\p^0.5/17); } 
  \draw (2.86,25.71) node[above]{Livakovic};
  \draw (11.11,25.00) node[above,xshift=12]{Donnarumma};
  \draw (0,21.62) node[above]{Bossut};
  \draw (14.71,14.71) node[above]{\scriptsize Sean};
  \draw (14.71,14.71) node[below]{\scriptsize O'Neill};
  \draw (5.13,5.13) node[below]{\scriptsize Matz Sels};
  \end{tikzpicture}
  \label{top50 goalies in-game}
}
\qquad
\subfloat[Top10 by percentage of saved penalties]
{\begin{tabular}[b]{lrrrr}
goalkeeper	& \llap{penalties} & goal & save & miss  \\ \hline
Dominik Livakovic& 35 & 71.43\% & 25.71\% & 2.86\% \\ 
Gianluigi Donnarumma&36& 63.89\%& 25.00\% &11.11\% \\ 
Mike Maignan	& 37  & 67.57\% & 24.32\% & 8.11\% \\ 
Yuriy Pankiv	& 38  & 71.05\% & 23.68\% & 5.26\% \\ 
Aleksandr Belenov& 41 & 68.29\% & 21.95\% & 9.76\% \\ 
Baptiste Reynet	& 41  & 75.61\% & 21.95\% & 2.44\% \\ 
Sammy Bossut    & 37  & 78.38\% & 21.62\% & -- \\ 
Wojciech Szcz{\c e}sny&56&73.21\%&21.42\% & 5.36\% \\ 
Du{\v s}an Kuciak& 47  & 72.34\% & 21.28\% & 6.38\% \\ 
Marafona	& 38  & 76.32\% & 21.05\% & 2.63\% \\ 
\end{tabular}
\label{top10 goalies in-game}
}}
\caption{Goalkeepers most frequently facing in-game penalties in dataset~D3.
There are 51~goalkeepers who faced 34~or more in-game penalties,
and dot size in the scatterplot on the left varies with that number.
The goal rate (percentage neither saved nor missed)
is constant along diagonals
with an overall average of~$82\%$ for in-game penalties.}
\label{top goalies in-game}
\end{figure}

\begin{figure}[t!]
\edef\goaliesshootout{%
Unai Simon/22/59.0909090909091/36.3636363636364/4.54545454545455,
Jason Steele/18/61.1111111111111/33.3333333333333/5.55555555555556,
Nestoras Gekas/18/50/33.3333333333333/16.6666666666667,
Egor Shamov/22/59.0909090909091/31.8181818181818/9.09090909090909,
Stanislav Antipin/29/62.0689655172414/31.0344827586207/6.89655172413793,
Beto/23/60.8695652173913/30.4347826086957/8.69565217391304,
Andrey Ryzhikov/25/60/28/12,
Vaclav Hladky/18/55.5555555555556/27.7777777777778/16.6666666666667,
Marvin Schwabe/19/73.6842105263158/26.3157894736842/0,
Gianluigi Donnarumma/23/60.8695652173913/26.0869565217391/13.0434782608696,
Eldin Jakupovic/24/66.6666666666667/25/8.33333333333333,
Teitur Gestsson/26/76.9230769230769/23.0769230769231/0,
Lubomir Belko/22/68.1818181818182/22.7272727272727/9.09090909090909,
Caoimhin Kelleher/27/74.0740740740741/22.2222222222222/3.7037037037037,
Danny Ward/18/61.1111111111111/22.2222222222222/16.6666666666667,
Ivan Rehak/18/72.2222222222222/22.2222222222222/5.55555555555556,
John Ruddy/18/61.1111111111111/22.2222222222222/16.6666666666667,
Paolo Branduani/18/66.6666666666667/22.2222222222222/11.1111111111111,
Alexander Meyer/23/69.5652173913043/21.7391304347826/8.69565217391304,
Lee Nicholls/23/65.2173913043478/21.7391304347826/13.0434782608696,
Simon Mignolet/24/75/20.8333333333333/4.16666666666667,
Dominik Livakovic/20/70/20/10,
Jordan Pickford/20/75/20/5,
Anthony Martin/26/80.7692307692308/19.2307692307692/0,
Bojan Zogovic/26/80.7692307692308/19.2307692307692/0,
Artur Nigmatullin/21/76.1904761904762/19.047619047619/4.76190476190476,
Lukas Hradecky/21/61.9047619047619/19.047619047619/19.047619047619,
Hobie Verhulst/22/72.7272727272727/18.1818181818182/9.09090909090909,
Timo Horn/22/77.2727272727273/18.1818181818182/4.54545454545455,
William Dutoit/22/68.1818181818182/18.1818181818182/13.6363636363636,
Kepa Arrizabalaga/29/68.9655172413793/17.2413793103448/13.7931034482759,
Leandro Chichizola/18/61.1111111111111/16.6666666666667/22.2222222222222,
Scott Carson/24/83.3333333333333/16.6666666666667/0,
Mark Gillespie/31/77.4193548387097/16.1290322580645/6.45161290322581,
Ioannis Gelios/19/78.9473684210526/15.7894736842105/5.26315789473684,
Géza Turi/20/80/15/5,
Manuel Neuer/20/70/15/15,
Nathan Vaughan/20/80/15/5,
Nikolay Moskalenko/20/70/15/15,
Sergio Romero/20/80/15/5,
Anton Antipov/21/71.4285714285714/14.2857142857143/14.2857142857143,
Aleksandr Kobzev/31/77.4193548387097/12.9032258064516/9.67741935483871,
Pavao Pervan/24/75/12.5/12.5,
Yann Sommer/32/81.25/12.5/6.25,
Artur Anisimov/26/76.9230769230769/11.5384615384615/11.5384615384615,
Rui Patricio/18/83.3333333333333/11.1111111111111/5.55555555555556,
Simon Thomas/18/72.2222222222222/11.1111111111111/16.6666666666667,
Edouard Mendy/37/81.0810810810811/10.8108108108108/8.10810810810811,
Joel Robles/19/84.2105263157895/10.5263157894737/5.26315789473684,
Angus Gunn/30/86.6666666666667/10/3.33333333333333,
Anuar Sapargaliev/20/85/10/5,
Aaron Chapman/25/92/8/0,
Anton Sytnykov/20/75/0/25,
David De Gea/18/94.4444444444444/0/5.55555555555556,
Mirko Salvi/18/77.7777777777778/0/22.2222222222222}
\centerline{\subfloat[percentage saved vs.\ missed]
 {\begin{tikzpicture}[scale=0.25]
  \begin{scope}
    \clip (0,0) rectangle (25,37);
    \foreach \x in {5,10,15,20,30,35,40} {\draw[gray!50] (\x,0) -- (0,\x);}
    \draw[blue,dashed] (25,0) -- node[below,near start,sloped]{$75\%$} (0,25);
  \end{scope}
  \draw[gray] (-1,0) node[green,left]{$0\%$} coordinate (B)
    -- node[orange,above,xshift=10]{percentage missed} (25,0);
  \draw[gray] (0,0) -- node[green,above,sloped,near start,yshift=5,xshift=-8]{percentage saved} (0,37);
  \foreach \x in {0,5,15,20,25} {\draw[gray] (\x,0) -- +(0,-1) node[orange,below]{$\x\%$};}
  \draw[orange,dashed] (10,-1) node[orange,below]{$10\%$} -- +(0,38);
  \draw[green,dashed] (-1,15) node[left]{$15\%$} -- +(26,0);
  \draw[green] (-1,23) node[left]{$23\%$} -- +(26,0) node[above left]{\tiny Top10};
  \foreach \n/\p/\r/\s/\m in \goaliesshootout { \fill (\m,\s) circle (\p^0.5/17); } 
  \draw (4.55,36.36) node[right]{Sim\'on};
  \draw (16.67,33.33) node[right]{Gekas};
  \draw (16.67,27.78) node[right]{Hladk\'y};
  \draw (0,26.32) node[right]{Schw\"abe};
  \draw (22.22,16.67) node[above]{\scriptsize Leandro} node[below]{\scriptsize Chichizola};
  \draw (25,0) node[left,rotate=-45,yshift=4]{\scriptsize Anton Sytnykov};
  \draw (5.56,0) node[left,rotate=-45,yshift=2]{\scriptsize David De Gea};
  \end{tikzpicture}
  \label{top50 goalies shootout}
}
\quad
\subfloat[Top10 by percentage of penalties saved]
{\begin{tabular}[b]{lrrrr}
goalkeeper	& \llap{penalties} & goal & save & miss  \\ \hline
Unai Sim\'on	& 22 & 59.09\% & 36.36\% & 4.54\% \\ 
Jason Steele	& 18 & 61.11\% & 33.33\% & 5.56\% \\ 
Nestoras Gekas	& 18 & 50.00\% & 33.33\% &16.67\% \\ 
Egor Shamov	& 22 & 59.09\% & 31.81\% & 9.09\% \\ 
Stanislav Antipin&29 & 62.07\% & 31.03\% & 6.90\% \\ 
Beto		& 23 & 60.87\% & 30.43\% & 8.70\% \\ 
Andrey Ryzhikov	& 25 & 60.00\% & 28.00\% &12.00\%  \\ 
V\'aclav Hladk\'y	& 18 & 55.56\% & 27.78\% & 16.67\% \\ 
Marvin Schw\"abe	& 19 & 73.68\% & 26.32\% & -- \\ 
Gianluigi Donnarumma&23&60.87\%& 26.09\% &13.04\% \\ 
Eldin Jakupovi\'c	& 24 & 66.67\% & 25.00\% & 8.33\% \\ 
Teitur Gestsson	& 26 & 76.92\% & 23.08\% & -- \\ 
\end{tabular}
\label{top10 goalies shootout}
}}
\caption{Goalkeepers most frequently facing shootout penalties in dataset~D3.
There are 55~goalkeepers who faced 18~or more shootout penalties,
and dot size in the scatterplot on the left varies with that number.
The overall average goal rate is~$75\%$ for shootout penalties.}
\label{top goalies shootout}
\end{figure}

\subsubsection{Goalkeepers}

The situation is qualitatively and quantitatively different for goalkeepers. 
Because of a routine distinction between
penalties saved (for which goalkeepers are praised)
and penalties missed (for which kickers are blamed) 
the analysis below is based on 
three-way outcome dataset~D3.
During a match goalkeepers are almost never changed or substituted 
because of an impending penalty, and very rarely so before a shootout.
Consequently, they have many more involvements in penalty situations
so that we can afford to distinguish
between in-game (\Cref{top goalies in-game})
and shootout penalties (\Cref{top goalies shootout}).

A threshold of 34~yields the 51~goalkeepers facing
the highest numbers of in-game penalties (\Cref{top goalies in-game}).
Despite the argument that
better goalkeepers make penalty kickers miss more often,
we find no correlation between the two rates in general.
Goalkeepers are therefore ranked by their save rate
in \Cref{top10 goalies in-game}.
Their collective save rate is 16.02\% compared to 13.08\% for all others.

While Livakovic has the highest rate of saved in-game penalties, 
there are three others (Donnarumma, Maignan, and Belenov) 
who conceded less than $70\%$
because they not only save a high fraction of penalties 
but kickers also missed at high rates.
The majority of frequent goalkeepers
has a lower rate of goals against than the average.
No one has faced more in-game penalties than Alban Lafont,
who saved eight out of 58~penalties.

The same analysis is repeated for goalkeepers
facing the highest number of penalties in the shootout condition.
To obtain a list of 55~goalkeepers,
the threshold had to be lowered to 18~shootout penalties,
Fewer observations are one of the reasons
the range of percentages is wider than for in-game penalties. 

There is a remarkable line of a dozen goalkeepers
who conceded around $60\%$ or fewer penalties in shootouts.
Their share of saved and missed penalties varies widely
from Unai Sim\'on to Leandro Chichizola.
Among frequently involved goalkeepers, 
Anton Sytnykov stands out with an average rate of conceded goals
exclusively due to misses, i.e., without saving a single one of them.
David De~Gea was not quite so lucky, 
conceding all eleven~penalties in the 2021 Europa League final and missing the deciding penalty himself. The 12:11 victory against Manchester United led to the first major trophy for Villareal FC. A single miss against David De~Gea occurred in the 2023 FA Cup semifinal against Brighton \& Hove Albion Football Club.

\subsubsection{Teams taking penalties}

Finally, we look at the collective performance of teams
during matches, where the choice of kicker is wide open,
and in shootouts,
where selection and preparation feature more prominently.

We consider only club teams,
because penalties for international sides
are even fewer and farther apart.

The thresholds used to obtain at least 50~teams frequently at the mark
are~67 for in-game and~25 for shootout penalties.
This is in line with the aggregate numbers indicating
that in-game penalties are twice as frequent as shootout penalties.

\begin{figure}[t!]
\edef\teamsingame{%
Fc Flora Tallinn/67/89.5522388059701/10.4477611940298/0,
Rsc Anderlecht/67/88.0597014925373/8.95522388059701/2.98507462686567,
Vitória Guimarães Sc/71/87.3239436619718/8.45070422535211/4.22535211267606,
Bayern Munich/76/86.8421052631579/7.89473684210526/5.26315789473684,
Lechia Gdansk/75/86.6666666666667/6.66666666666667/6.66666666666667,
Sl Benfica/86/86.046511627907/10.4651162790698/3.48837209302326,
Fci Levadia/70/85.7142857142857/11.4285714285714/2.85714285714286,
As Roma/82/85.3658536585366/9.75609756097561/4.8780487804878,
Juventus Fc/95/85.2631578947368/8.42105263157895/6.31578947368421,
Acf Fiorentina/80/85/11.25/3.75,
Liverpool Fc/73/84.9315068493151/9.58904109589041/5.47945205479452,
As Monaco/78/84.6153846153846/12.8205128205128/2.56410256410256,
Olympique Lyon/83/84.3373493975904/9.63855421686747/6.02409638554217,
Paok Thessaloniki/76/84.2105263157895/11.8421052631579/3.94736842105263,
Gnk Dinamo Zagreb/88/84.0909090909091/11.3636363636364/4.54545454545455,
Losc Lille/69/84.0579710144928/14.4927536231884/1.44927536231884,
Olympiacos Piraeus/73/83.5616438356164/13.6986301369863/2.73972602739726,
Paris Saint-Germain/100/83/14/3,
Celtic Fc/76/82.8947368421053/14.4736842105263/2.63157894736842,
Panathinaikos Athens/70/82.8571428571429/12.8571428571429/4.28571428571429,
Besiktas Jk/81/82.7160493827161/9.87654320987654/7.40740740740741,
Sporting Cp/97/82.4742268041237/12.3711340206186/5.15463917525773,
Az Alkmaar/78/82.051282051282/11.5384615384615/6.41025641025641,
Hnk Hajduk Split/89/82.0224719101124/8.98876404494382/8.98876404494382,
Fc Basel 1893/83/81.9277108433735/12.0481927710843/6.02409638554217,
Hnk Rijeka/87/81.6091954022989/14.9425287356322/3.44827586206897,
Red Star Belgrade/102/81.3725490196078/10.7843137254902/7.84313725490196,
Chelsea Fc/96/81.25/14.5833333333333/4.16666666666667,
Manchester United/95/81.0526315789474/12.6315789473684/6.31578947368421,
Fc Luzern/73/80.8219178082192/15.0684931506849/4.10958904109589,
Ajax Amsterdam/81/80.2469135802469/13.5802469135802/6.17283950617284,
Dynamo Kyiv/81/80.2469135802469/11.1111111111111/8.64197530864197,
Bsc Young Boys/105/80/14.2857142857143/5.71428571428571,
Dundalk Fc/70/80/14.2857142857143/5.71428571428571,
Fc Zürich/75/80/12/8,
Ssc Napoli/90/80/11.1111111111111/8.88888888888889,
Ac Milan/74/79.7297297297297/13.5135135135135/6.75675675675676,
Red Bull Salzburg/93/79.5698924731183/11.8279569892473/8.60215053763441,
Ss Lazio/93/79.5698924731183/15.0537634408602/5.37634408602151,
Fk Partizan Belgrade/73/79.4520547945205/16.4383561643836/4.10958904109589,
Shakhtar Donetsk/82/78.0487804878049/18.2926829268293/3.65853658536585,
Ferencvárosi Tc/67/77.6119402985075/16.4179104477612/5.97014925373134,
Psv Eindhoven/71/77.4647887323944/18.3098591549296/4.22535211267606,
Trabzonspor/75/76/17.3333333333333/6.66666666666667,
Fc Porto/90/75.5555555555556/17.7777777777778/6.66666666666667,
Nk Osijek/69/75.3623188405797/15.9420289855072/8.69565217391304,
Fc St. Gallen 1879/72/75/16.6666666666667/8.33333333333333,
Zenit St. Petersburg/67/74.6268656716418/19.4029850746269/5.97014925373134,
Feyenoord Rotterdam/74/74.3243243243243/20.2702702702703/5.40540540540541,
Manchester City/94/72.3404255319149/17.0212765957447/10.6382978723404,
Slaven Belupo Koprivnica/67/71.6417910447761/20.8955223880597/7.46268656716418,
Zorya Lugansk/67/70.1492537313433/17.910447761194/11.9402985074627}
\centerline{\subfloat[percentage saved vs.\ missed]
 {\begin{tikzpicture}[scale=0.25]
  \begin{scope}
    \clip (0,0) rectangle (15,22);
    \foreach \x in {10,15,20,25} {\draw[gray!50] (\x,0) -- (0,\x);}
    \draw[blue,dashed] (18,0) -- node[below,near start,sloped]{$82\%$} (0,18);
  \end{scope}
  \draw[gray] (-1,0) node[green,left]{$0\%$} coordinate (B)
    -- node[orange,above]{percentage missed} (15,0);
  \draw[gray] (0,0) -- node[green,above,sloped,near start,yshift=5,xshift=5]{percentage saved} (0,22);
  \foreach \x in {0,10,15} {\draw[gray] (\x,0) -- +(0,-1) node[orange,below]{$\x\%$};}
  \draw[orange,dashed] (5,-1) node[below]{$5\%$} --  (5,22);
  \draw[green,dashed] (-1,13) node[left]{$13\%$} --  (15,13);
  \draw[gray] (0,20) -- +(-1,0) node[green,left]{$20\%$};
  \foreach \n/\p/\r/\s/\m in \teamsingame { \fill (\m,\s) circle (\p^0.5/25); } 
  \draw (0,10.45) node[right,rotate=-45]{Tallinn};
  \draw (7.46,20.90) node[right]{\scriptsize NK Slaven Belupo};
  \draw (11.94,17.91) node[above,xshift=8]{\scriptsize Sorya Luhansk};
  \end{tikzpicture}
  \label{top50 teams in-game}
}
\qquad
\subfloat[Top10 by goal rate]
{\begin{tabular}[b]{lrrrr}
club team & penalties & goal & save & miss  \\ \hline
FC Flora Tallinn& 67 & 89.55\% & 10.45\% &   -- \\ 
RSC Anderlecht	& 67 & 88.06\% &  8.96\% & 2.99\% \\ 
Vit\'oria Guimar{\~ a}es SC& 71 & 87.32\% &  8.45\% & 4.23\% \\ 
Bayern M\"unchen& 76 & 86.84\% &  7.89\% & 5.26\% \\ 
Lechia Gdansk 	& 75 & 86.67\% &  6.67\% & 6.67\% \\ 
SL Benfica 	& 86 & 86.04\% & 10.47\% & 3.49\% \\ 
FCI Levadia 	& 70 & 85.71\% & 11.43\% & 2.86\% \\ 
AS Roma 	& 82 & 85.37\% &  9.76\% & 4.88\% \\
Juventus FC 	& 95 & 85.26\% &  8.42\% & 6.32\% \\ 
ACF Fiorentina 	& 80 & 85.00\% & 11.25\% & 3.75\% \\ 
Liverpool FC    & 73 & 84.92\% &  9.59\% & 5.48\% \\
AS Monaco 	& 78 & 84.62\% & 12.82\% & 2.56\% \\ 
\end{tabular}
\label{top10 teams in-game}
}}
\caption{Most frequent teams (minimum of 67~in-game penalties).}
\label{top teams in-game}
\end{figure}

Unsurprisingly, the data in \Cref{top teams in-game}
shows that the most successful teams reduce both save and miss rates,
but that their larger number of collective penalties
reduces their advantage over average penalty taker rather quickly.
The combined conversion rate for the 52~most frequent teams is 81.16\%
and similar to the rate for all other teams of 82.07\%. 

\begin{figure}[b!]
\edef\teamsshootout{%
Sevilla Fc/25/92/8/0,
Fc Tosno (-2018)/25/88/12/0,
Stoke City/33/87.8787878787879/6.06060606060606/6.06060606060606,
Fk Khimki/32/87.5/9.375/3.125,
Derby County/39/87.1794871794872/10.2564102564103/2.56410256410256,
Újpest Fc/31/87.0967741935484/6.45161290322581/6.45161290322581,
Manchester United/43/86.046511627907/9.30232558139535/4.65116279069767,
Sv Sandhausen/28/85.7142857142857/7.14285714285714/7.14285714285714,
Fk Tyumen/26/84.6153846153846/7.69230769230769/7.69230769230769,
Hull City/48/83.3333333333333/10.4166666666667/6.25,
Lask/30/83.3333333333333/3.33333333333333/13.3333333333333,
Queens Park Rangers/41/82.9268292682927/9.75609756097561/7.31707317073171,
Sokol Saratov/35/82.8571428571429/11.4285714285714/5.71428571428571,
Chelsea Fc/64/82.8125/10.9375/6.25,
Gillingham Fc/34/82.3529411764706/14.7058823529412/2.94117647058824,
Eintracht Frankfurt/28/82.1428571428571/17.8571428571429/0,
Krc Genk/28/82.1428571428571/14.2857142857143/3.57142857142857,
Liverpool Fc/60/81.6666666666667/11.6666666666667/6.66666666666667,
Iraklis Thessaloniki/27/81.4814814814815/3.7037037037037/14.8148148148148,
Spartak Trnava/27/81.4814814814815/7.40740740740741/11.1111111111111,
Southampton Fc/31/80.6451612903226/9.67741935483871/9.67741935483871,
Ac Sparta Prague/35/80/8.57142857142857/11.4285714285714,
Tsv Hartberg/25/80/16/4,
Spartak Nalchik/34/79.4117647058823/17.6470588235294/2.94117647058824,
Wycombe Wanderers/28/78.5714285714286/10.7142857142857/10.7142857142857,
Ssv Jahn Regensburg/27/77.7777777777778/18.5185185185185/3.7037037037037,
Tom Tomsk/27/77.7777777777778/11.1111111111111/11.1111111111111,
Everton Fc/31/77.4193548387097/16.1290322580645/6.45161290322581,
Carlisle United/26/76.9230769230769/23.0769230769231/0,
Fc Slovan Liberec/26/76.9230769230769/15.3846153846154/7.69230769230769,
Levski Sofia/30/76.6666666666667/16.6666666666667/6.66666666666667,
Accrington Stanley/42/76.1904761904762/19.047619047619/4.76190476190476,
Pogon Szczecin/25/76/16/8,
1.Fc Nuremberg/29/75.8620689655172/20.6896551724138/3.44827586206897,
Dynamo Kyiv/27/74.0740740740741/18.5185185185185/7.40740740740741,
Fc Luzern/27/74.0740740740741/22.2222222222222/3.7037037037037,
Leeds United/27/74.0740740740741/7.40740740740741/18.5185185185185,
Leicester City/45/73.3333333333333/17.7777777777778/8.88888888888889,
1. Fc Köln/26/73.0769230769231/11.5384615384615/15.3846153846154,
Fk Chita/25/72/20/8,
Volga Uljanovsk/25/72/20/8,
Ingulets Petrove/32/71.875/18.75/9.375,
Stade Rennais Fc/42/71.4285714285714/21.4285714285714/7.14285714285714,
Stade Brestois 29/27/70.3703703703704/18.5185185185185/11.1111111111111,
Msk Zilina/33/69.6969696969697/24.2424242424242/6.06060606060606,
Hb Tórshavn/26/69.2307692307692/23.0769230769231/7.69230769230769,
Real Betis Balompié/26/69.2307692307692/19.2307692307692/11.5384615384615,
Tottenham Hotspur/34/67.6470588235294/20.5882352941176/11.7647058823529,
Bradford City/33/66.6666666666667/27.2727272727273/6.06060606060606,
Norwich City/27/66.6666666666667/22.2222222222222/11.1111111111111,
Amkar Perm/44/65.9090909090909/25/9.09090909090909,
Brighton & Hove Albion/31/64.5161290322581/12.9032258064516/22.5806451612903,
Cfr Cluj/25/64/24/12,
Luki-Energia Velikie Luki/27/62.962962962963/14.8148148148148/22.2222222222222,
Sv Mattersburg/31/61.2903225806452/19.3548387096774/19.3548387096774,
Ska Khabarovsk/25/60/16/24}
\centerline{\subfloat[percentage saved vs.\ missed]
 {\begin{tikzpicture}[scale=0.25]
  \begin{scope}
    \clip (0,0) rectangle (25,28);
    \foreach \x in {10,15,20,30,35} {\draw[gray!50] (\x,0) -- (0,\x);}
    \draw[blue,dashed] (25,0) -- node[below,very near start,sloped]{$75\%$} (0,25);
  \end{scope}
  \draw[gray] (-1,0) node[green,left]{$0\%$} coordinate (B)
    -- node[orange,above,near start,xshift=8]{percentage missed} (25,0);
  \draw[gray] (0,0) -- node[green,above,sloped,near start,yshift=5,xshift=5]{percentage saved} (0,28);
  \foreach \x in {0,5,15,20,25} {\draw[gray] (\x,0) -- +(0,-1) node[orange,below]{$\x\%$};}
  \draw[orange,dashed] (10,-1) node[below]{$10\%$} --  (10,28);
  \draw[green,dashed] (-1,15) node[left]{$15\%$} --  (25,15);
  \draw[gray] (0,20) -- +(-1,0) node[green,left]{$20\%$};
  \foreach \n/\p/\r/\s/\m in \teamsshootout { \fill (\m,\s) circle (\p^0.5/25); } 
  \draw (0,8) node[right,rotate=-45]{Sevilla};
  \draw (0,23.08) node[above,xshift=2]{\scriptsize Carlisle United};
  \draw (18.52,7.40) node[above]{\scriptsize Leeds United};
  \draw (6.06,27.27) node[left]{\scriptsize Bradford City};
  \draw (19.35,19.35) node[above]{\scriptsize SV Mattersburg};
  \draw (24,16) node[above left]{\scriptsize SKA Khabarovsk};
  \end{tikzpicture}
  \label{top50 teams shootout}
}
\qquad
\subfloat[Top10 by goal rate]
{\begin{tabular}[b]{lrrrr}
club team & \llap{penalties} & goal & save & miss  \\ \hline
Sevilla FC	& 25 & 92.00\% & 8.00\% &  --    \\ 
FK Tosno        & 25 & 88.00\% &12.00\% &   --   \\ 
Stoke City	& 33 & 87.88\% & 6.06\% & 6.06\% \\ 
FK Khimki	& 32 & 87.50\% & 9.38\% & 3.13\% \\ 
Derby County	& 39 & 87.18\% &10.26\% & 2.56\% \\ 
\'Ujpest FC 	& 31 & 87.10\% & 6.45\% & 6.45\% \\ 
Manchester United& 43& 86.05\% & 9.30\% & 4.65\% \\ 
SV Sandhausen	& 28 & 85.71\% & 7.14\% & 7.14\% \\ 
FK Tyumen	& 26 & 84.62\% & 7.69\% & 7.69\% \\ 
Hull City 	& 48 & 83.33\% &10.42\% & 6.25\% \\ 
\end{tabular}
\label{top10 teams shootout}
}}
\caption{Most frequent teams (25~or more shootout penalties).}
\label{top teams shootout}
\end{figure}

For shootout penalties shown in \Cref{top teams shootout},
smaller numbers lead to more variation in conversion rates.
While teams with many shootout penalties 
do not necessarily score at a higher rate,
few of them average more misses than the overall average.
Teams such as Carlisle United and Leeds United
exhibit average goal rates in very different ways.

Clearly, it helps not to miss any penalty during a shootout.
When, in addition, a team such as Sevilla~FC has only two of 25~penalties saved,
chances are they do well in cup competitions.%
\footnote{The UEFA Super Cup 2023 
in which Sevilla~FC lost the penalty shootout against Manchester City
by missing the last penalty after nine goals 
was played after our cut-off date 
and in a single-match competition we do not cover.}

The two teams with the highest number of shootout penalties in~D3
are Chelsea~FC and Liverpool~FC,
taking~64 and~60 penalties, respectively,
whereas all other teams took less than~50.
These two are represented by the largest dots
in the lower left quadrant of \Cref{top50 teams shootout}
and still accomplished goal rates just below the Top10.

We next turn to possible explanations for the empirical finding that 
penalties are converted less frequently in shootouts than during regulation.

\subsection{Player selection}\label{sec:selection}

An obvious difference between
penalties during regulation and in shooutouts is that 
penalty kickers can be chosen freely during a match,
but at least five different ones are to be nominated for a shootout.

Assuming that in-game penalties are generally taken by the most reliable kickers,
we should indeed expect that shootout penalities
have a lower goal percentage,
because they involve players who are not likely to take a penalty
absent the selection constraints~\citep{arrondel:2019}. 

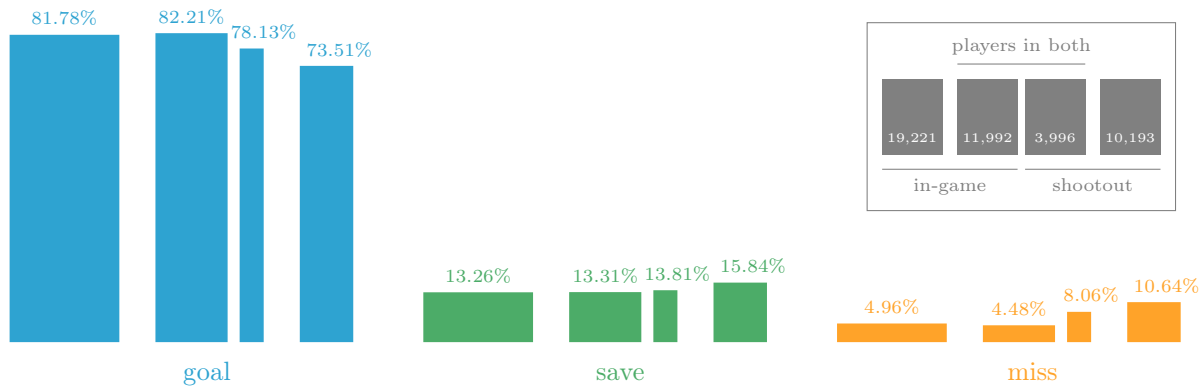
\begin{figure}
\centering
\centerline{\begin{tikzpicture}
\def\barscale{20}
\def\nmatchonly{1.8235}
\def\nmatchboth{1.1992}
\def\nshootoutboth{0.3996}
\def\nshootoutonly{0.8882}
\tikzset{pics/fbars/.style n args={5}{code={%
\fill[xscale=0.8]
  (0,0) rectangle ++(\nmatchonly,#1/\barscale) +(-\nmatchonly/2,0) node[above]{\scriptsize$#1\%$}
  (0.6+\nmatchonly,0) rectangle ++(\nmatchboth,#2/\barscale) +(-\nmatchboth/2,0) node[above]{\scriptsize$#2\%$}
  (0.8+\nmatchonly+\nmatchboth,-0.5) node[below,anchor=base east]{#5}
  (0.8+\nmatchonly+\nmatchboth,0) rectangle ++(\nshootoutboth,#3/\barscale) node[above]{\scriptsize$#3\%$}
  (1.4+\nmatchonly+\nmatchboth+\nshootoutboth,0) rectangle ++(\nshootoutonly,#4/\barscale) +(-\nshootoutonly,0) node[above right,xshift=-1]{\scriptsize$#4\%$}
;
}}}
\path pic[blue]{fbars={81.78}{82.21}{78.13}{73.51}{goal}}
  (5.5,0) pic[green]{fbars={13.26}{13.31}{13.81}{15.84}{save}}
  (11,0) pic[orange]{fbars={4.96}{4.48}{8.06}{10.64}{miss}};
\draw[gray] (12,2.5) coordinate (L)++(-0.6,-0.75) rectangle +(4.1,2.5);
\foreach \x/\n in {0/{19,221},1/{11,992},1.9/{3,996},2.9/{10,193}}
 {
  \fill[gray] (L)++(\x-0.4,0) rectangle +(0.8,1); 
  \path[white] (L)++(\x,0) node[above]{\tiny \n}; 
 }
\draw[gray] (L)
  ++(-0.4,-0.2) -- node[below]{\scriptsize in-game} ++(1.8,0)
  ++(0.1,0) -- node[below]{\scriptsize shootout} ++(1.8,0)
  (L)
  ++(0.6,1.2) -- node[above]{\scriptsize players in both} ++(1.7,0);
\end{tikzpicture}}
\caption{
Relative frequencies of outcomes in~D3
split by condition players appear in.
Only 2,855~players have taken both in-game and shootout penalties.
Bar widths are proportional to the number of penalties.}
\label{outcome appearance}
\end{figure}

We find that this is only part of the explanation,
because even those who are taking the penalties awarded during a match
see their success rate drop in shootouts.
This is illustrated by \Cref{outcome appearance},
where two categories of players are compared:
those who have taken penalties both during matches and in shootouts
and those who have only taken penalties during shootouts.
Since many of the first group of players are selected repeatedly, 
the ratio of in-game to shootout penalties is more skewed than overall.
Those taking a penalty during a match are chosen deliberately, 
so that their in-game goal rate corresponds to the overall average as expected. 
Although only by four percentage points,
even these players perform worse in shootouts,
and their decreased success rate is almost entirely due to an increased rate of missing.
For the second group of players,
both save and miss rate are at least two percentage points higher
than in the first group. 

While the combined four and a half percentage point difference 
may reflect a quality gap between the two groups,
the drop in performance within the first group  
is attributed to the shootout condition.

To test this further, we ran a binary logistic regression on penalty success.
The outcome variable is goal or no goal, 
and a number of binary fixed-effects are included as listed in \Cref{logreg}.
The reference level (all binary variables zero) is
an in-game penalty in a national league,
taken by a player of the away team
with the experience of one in-game and one shootout penalty.
A kicker is considered to appear at the mark repeatedly, 
if he has taken penalties in both settings
with more than one penalty in at least one of them.
We included seasons by the year they start in
and ignored that in a comparatively small number of cases 
such as international tournaments and international cup finals
in which a team labeled home team is not actually playing at their home ground.

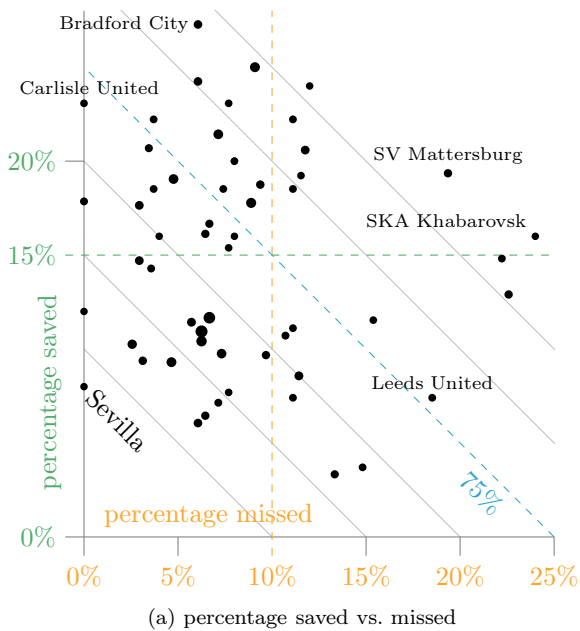
\begin{table}
\caption{Binary logistic regression of penalty outcomes in D2.
All variables are coded as binary variables except for the season.}
\begin{tabular}{lrrrrr}
 variable& estimate & exp$^\text{estimate}$ & std.~error & z value & $\mathbf{p}^\dagger$ \\ \hline
 (intercept) & -10.242 & 0.0000 & 7.4747 & -1.3702 & 0.3531\\ 
  shootout\,$^{***}$ & -0.3867 & 0.6793 & 0.0498 & -7.7608 & 0.0000 \\ 
  \hline
  \multicolumn{6}{l}{players}\\
  \hline
  repeated appearance\,$^{***}$ & 0.3056 & 1.3574 & 0.0583 & 5.2378 & 0.0000\\ 
  only in-game\,$^{***}$ & 0.2232 & 1.2501 & 0.0610 & 3.6578 & 0.0018\\ 
  only shootout & -0.0321 & 0.9684 & 0.0600 & -0.5351 & 0.5926\\ 
  \hline
  \multicolumn{6}{l}{competition}\\
  \hline
  national cup\,$^{***}$ & -0.2352 & 0.7904 & 0.0364 & -6.4552 & 0.0000\\ 
  international cup & -0.1808 & 0.8346 & 0.0794 & -2.2771 & 0.1139\\ 
  international selections\,$^{***}$ & -0.3663 & 0.6933 & 0.1111 & -3.2959 & 0.0059\\ 
  \hline
  \multicolumn{6}{l}{environment}\\
  \hline
  home team & -0.0381 & 0.9626 & 0.0220 & -1.7335 & 0.3320\\ 
  season & 0.0058 & 1.0058 & 0.0037 & 1.5645 & 0.3531\\ [2ex]

\multicolumn{3}{l}{$^{*}p<0.1$\qquad $^{**}p<0.05$\qquad $^{***}p<0.01$} &
\multicolumn{3}{r}{$^\dagger$Bonferroni-Holm adjusted}
\\ 
\end{tabular}
\label{logreg}
\end{table} 

Clearly, the shootout condition stands out as the most important factor 
decreasing penalty success rates. 
If players taking penalties repeatedly
were not significantly more successful on average, 
their selection would reflect poor decision-making.
Players having taken only in-game penalties have been picked because of their quality
and may not have been involved in shootouts.
The effect is positive and significant. 
Players only taking shootout penalties, however,
have not been picked during regulation,
likely because there is at least one more reliable kicker on the team.
Although indeed negative, this effect is thus smaller and not significant.

Other significant effects are related to the type of competition,
but confounded with shootouts
because they feature only in club cup competitions and international-side tournaments 
and are an indication of competitive balance and high stakes.
Interestingly, we find no significant effect for home teams.

\subsection{Fatigue?}\label{sec:fatigue}

With the above evidence that it is harder to convert penalties in a shootout,
we can turn to possible explanations.
A commonly offered one
is that shootouts take place at the end of a match~--
after regulation and, usually, extra time. 
Many on-field players will have played for 120~minutes by then,
and fatigue may reduce their ability to score.
The limited empirical evidence that there is, however,
does not support this reasoning~\citep{jordet:2007,jamil:2020}.

The argument would be more plausible
if we were to observe declining conversion rates
over the duration of matches. 
In \Cref{fig:timeseries},
we plot the number of penalties in each minute
against the corresponding conversion rate
averaged over a surrounding five-minute intervals
to smooth the curve and remove timing imprecisions.

\def\ratesHone{%
83.8212634822804/3.85208012326656/12.326656394453/154/3,
82.9638273045508/3.96732788798133/13.0688448074679/169/4,
82.0043103448276/3.87931034482759/14.1163793103448/195/5,
81.4412635735439/4.04738400789733/14.5113524185587/219/6,
81.5740740740741/3.7962962962963/14.6296296296296/191/7,
81.9148936170213/4.34397163120567/13.741134751773/239/8,
82.3784722222222/4.25347222222222/13.3680555555556/236/9,
81.8032786885246/4.83606557377049/13.3606557377049/243/10,
82.2314049586777/4.87603305785124/12.8925619834711/243/11,
82.1548821548822/4.71380471380471/13.1313131313131/259/12,
81.0463121783877/4.80274442538593/14.1509433962264/229/13,
81.5436241610738/4.86577181208054/13.5906040268456/214/14,
82.8035859820701/4.48247758761206/12.7139364303178/221/15,
83.5880933226066/3.86162510056315/12.5502815768302/269/16,
84.1426403641882/4.17298937784522/11.6843702579666/294/17,
85/3.84057971014493/11.1594202898551/245/18,
83.6428571428571/4.35714285714286/12/289/19,
82.9127613554434/4.39798125450613/12.6892573900505/283/20,
82.1149751596877/4.61320085166785/13.2718239886444/289/21,
81.4203730272597/4.73457675753228/13.845050215208/281/22,
81.5582558970693/4.93209435310936/13.5096497498213/267/23,
82.0440028388928/4.54222853087296/13.4137686302342/274/24,
81.6455696202532/4.92264416315049/13.4317862165963/288/25,
81.038062283737/5.32871972318339/13.6332179930796/299/26,
80.998613037448/5.33980582524272/13.6615811373093/294/27,
80.8275862068966/5.37931034482759/13.7931034482759/290/28,
80.6693989071038/5.46448087431694/13.8661202185792/271/29,
80.603152844414/5.14050719671008/14.2563399588759/296/30,
80.9078771695594/5.34045393858478/13.7516688918558/313/31,
81.635301752109/5.25632706035042/13.1083711875406/289/32,
81.8858560794045/5.27295285359801/12.8411910669975/329/33,
81.8407960199005/5.03731343283582/13.1218905472637/314/34,
82.6034063260341/5.17031630170316/12.2262773722628/367/35,
82.9518072289157/4.27710843373494/12.7710843373494/309/36,
82.5301204819277/3.97590361445783/13.4939759036145/325/37,
82.1273964131107/4.143475572047/13.7291280148423/345/38,
82.4529447480267/4.00728597449909/13.5397692774742/314/39,
82.8307692307692/3.63076923076923/13.5384615384615/324/40,
83.3644859813084/3.67601246105919/12.9595015576324/339/41,
83.252131546894/3.77588306942753/12.9719853836784/303/42,
83.5360554078621/3.64283283219517/12.8211117599428/325/43,
83.7326797825236/4.02311464176005/12.2442055757164/351/44,
83.4999315322741/4.25306935986439/12.2469991078615/336.582944603664/45,
83.1742243436754/4.71360381861575/12.1121718377088/329.472351798779/46
}
\def\ratesHtwo{%
81.5111111111111/5.86666666666667/12.6222222222222/289/48,
82.5/5.625/11.875/309/49,
82.8712261244609/5.23721503388786/11.8915588416513/318/50,
82.3668639053254/4.49704142011834/13.1360946745562/342/51,
82.9959514170041/4.16425679583574/12.8397917871602/365/52,
82.890365448505/4.31893687707641/12.7906976744186/356/53,
82.7180310326378/4.38737292669877/12.8945960406635/348/54,
82.1195652173913/4.72826086956522/13.1521739130435/395/55,
82.5107296137339/4.72103004291846/12.7682403433476/405/56,
82.6645264847512/4.38737292669877/12.94810058855/336/57,
81.5135135135135/4.7027027027027/13.7837837837838/380/58,
80.9898762654668/4.78065241844769/14.2294713160855/353/59,
81.5149802148106/4.18315432447711/14.3018654607123/376/60,
81.6694867456289/4.39932318104907/13.9311900733221/333/61,
81.4019219898248/4.74844544940644/13.8496325607688/327/62,
82.6599326599327/4.54545454545455/12.7946127946128/384/63,
82.6159865846842/4.58356623812186/12.800447177194/349/64,
81.8380743982494/5.14223194748359/13.019693654267/389/65,
82.0199778024417/5.1054384017758/12.8745837957825/340/66,
81.953642384106/5.29801324503311/12.7483443708609/366/67,
81.3703284258211/5.26613816534541/13.3635334088335/358/68,
81.2957746478873/5.07042253521127/13.6338028169014/359/69,
81.859410430839/4.59183673469388/13.5487528344671/343/70,
81.6462736373749/4.56062291434928/13.7931034482759/349/71,
81.4752252252252/4.5045045045045/14.0202702702703/355/72,
80.9684684684685/4.67342342342342/14.3581081081081/392/73,
81.5290178571429/4.85491071428571/13.6160714285714/337/74,
81.6750983698707/5.05902192242833/13.265879707701/343/75,
81.6372939169983/5.11654349061967/13.246162592382/365/76,
81.6338028169014/4.95774647887324/13.4084507042254/342/77,
82.6014452473596/4.66926070038911/12.7292940522512/372/78,
82.3397075365579/4.49943757030371/13.1608548931384/353/79,
82.1567537520845/4.55808782657032/13.2851584213452/367/80,
81.7780231916068/4.85919381557151/13.3627829928216/344/81,
82.1350762527233/4.68409586056645/13.1808278867102/363/82,
81.9520174482006/4.90730643402399/13.1406761177754/384/83,
81.9925412892914/4.90143846563665/13.1060202450719/378/84,
81.4873417721519/4.64135021097046/13.8713080168776/365/85,
82.4317086234601/4.23138725227638/13.3369041242635/387/86,
82.1615949632739/4.40713536201469/13.4312696747114/382/87,
81.6932580692222/4.53485920253341/13.7718827282444/355/88,
81.2686847440989/4.77665982115082/13.9546554347503/417/89,
81.2047973173818/5.29496261237742/13.5002400702408/396.010324738459/90,
80.5720035429011/5.62127095851617/13.8067254985828/423.597935052308/91,
80.151153540175/5.76770087509944/14.0811455847255/423.597935052308/92,
80.151153540175/5.76770087509944/14.0811455847255/423.597935052308/93
}
\def\ratesEone{%
89.4736842105263/5.26315789473684/5.26315789473684/295.706445993031/93,
90.4761904761905/4.76190476190476/4.76190476190476/349.471254355401/94,
91.304347826087/4.34782608695652/4.34782608695652/295.706445993031/95,
88.6363636363636/6.81818181818182/4.54545454545455/161.294425087108/96,
88.6363636363636/6.81818181818182/4.54545454545455/134.412020905923/97,
89.4736842105263/7.89473684210526/2.63157894736842/241.941637630662/98,
85.3658536585366/9.75609756097561/4.8780487804878/349.471254355401/99,
83.3333333333333/10.4166666666667/6.25/134.412020905923/100,
87.5/6.25/6.25/241.941637630662/101,
90.9090909090909/4.54545454545455/4.54545454545455/322.588850174216/102,
87.012987012987/3.8961038961039/9.09090909090909/241.941637630662/103
}
\def\ratesEtwo{%
78.7878787878788/12.1212121212121/9.09090909090909/241.941637630662/108,
81.8181818181818/9.09090909090909/9.09090909090909/268.824041811847/109,
84.7826086956522/6.52173913043478/8.69565217391304/161.294425087108/110,
88.8888888888889/2.22222222222222/8.88888888888889/376.353658536585/111,
95/0/5/188.176829268293/112,
94.8717948717949/0/5.12820512820513/215.059233449477/113,
94.1176470588235/0/5.88235294117647/134.412020905923/114,
92.1052631578947/0/7.89473684210526/134.412020905923/115,
90.2439024390244/0/9.75609756097561/241.941637630662/116,
88.6792452830189/0/11.3207547169811/295.706445993031/117,
88.75/0/11.25/295.706445993031/118
}
\begin{figure}
\centering
\begin{tikzpicture}[scale=0.095]
\def\bargap{17}
\def\barscale{50}
\def\offsetHtwo{5} \def\offsetEone{13} \def\offsetEtwo{15}
\draw[thin,gray]
  (-2,0) node[left]{$0\%$} -- ++(2,0)
  (1,0) -- +(44,0)
  (\offsetHtwo+46,0) -- +(44,0)
  (\offsetEtwo+122,0) -- +(2,0) node[right]{$0\%$};
\draw[thin,gray!50]
  (\offsetEone+91,0) -- +(14,0)
  (\offsetEtwo+106,0) -- +(14,0);
\foreach \t in {1,15,30,45}
 {\draw[thin,gray] (\t,0) -- +(0,-1) node[below]{\t};}
\foreach \t in {46,60,75,90}
 {\draw[thin,gray] (\offsetHtwo+\t,0) -- +(0,-1) node[below]{\t};}
\draw[thin,gray]
  (\offsetEone+91,0) -- +(0,-1) node[below]{91}
  (\offsetEone+105,0) -- ++(0,-1) +(-2,0) node[below]{105};
\draw[thin,gray]
  (\offsetEtwo+106,0) -- ++(0,-1) +(2,0) node[below]{106}
  (\offsetEtwo+120,0) -- ++(0,-1) +(-2,0) node[below]{120};
\draw[thin,gray]
  (0,0) -- ++(0,18) ++(0,8) -- (0,100-50) -- +(-2,0) node[left]{$100\%$}
  (\offsetEtwo+122,0) -- ++(0,18)
  ++(0,8) -- (\offsetEtwo+122,100-50) -- +(2,0) node[right]{$100\%$};
\draw[thin,gray,dotted]
  (0,18) -- +(0,8)
  (\offsetEtwo+122,18) -- +(0,8);
\draw[blue]
  (0,82-50) -- +(-2,0) node[left]{$82\%$}
  (\offsetEtwo+122,82-50) -- +(2,0) node[right]{$82\%$};
\draw[green]
  (0,13) -- +(-2,0) node[left]{$13\%$}
  (\offsetEtwo+122,13) -- +(2,0) node[right]{$13\%$};
\draw[orange]
  (0,5) -- +(-2,0) node[left]{$5\%$}
  (\offsetEtwo+122,5) -- +(2,0) node[right]{$5\%$};
\draw[thin,gray] 
  (-2,300/\barscale-\bargap) node[left]{$300$} -- +(95,0);
\draw[thin,gray!50] 
  (\offsetEone+91,269/\barscale-\bargap) -- +(34,0) node[right]{$10$};
\tikzset{pics/fbars/.style n args={1}{code={%
\fill ++(-0.5,0) foreach \f in {#1} {++(1,0) rectangle +(1,\f/\barscale)};
}}}
\tikzset{pics/rcurves/.style n args={6}{code={%
\path
  (#1+#2,#3-50) coordinate (G)  
  (#1+#2,#4) coordinate (S)
  (#1+#2,#5) coordinate (M);
\foreach \g/\m/\s/\n/\t in #6
 {
  \fill (#1+\t-0.5,-\bargap) rectangle +(1,\n/\barscale);
  \draw[blue] (G) -- (#1+\t,\g-50) coordinate (G);
  \draw[green] (S) -- (#1+\t,\s) coordinate (S);
  \draw[orange] (M) -- (#1+\t,\m) coordinate (M);
 }
}}}
\pic[very thick,gray,transform shape]
  {rcurves={0}{3}{83.82}{12.33}{3.85}{\ratesHone}};
\pic[gray,transform shape] at (0,-\bargap) {fbars={11,120}};
\pic[gray!75,transform shape] at (45,-\bargap) {fbars={329.47,329.47,329.47}};
\draw[orange] (30,2.5) node{misses};
\draw[green] (30,16) node{saves};
\draw[blue] (30,85-50) node{goals};
\pic[very thick,gray,transform shape]
  {rcurves={\offsetHtwo}{48}{81.51}{12.62}{5.87}{\ratesHtwo}};
\pic[gray,transform shape] at (\offsetHtwo+45,-\bargap) {fbars={27,182}};
\pic[gray!75,transform shape] at (\offsetHtwo+90,-\bargap) {fbars={423.6,423.6,423.6,423.6,423.6}};
\pic[very thin,gray!50,transform shape]
  {rcurves={\offsetEone}{93}{89.47}{5.25}{5.26}{\ratesEone}};
\pic[gray!50,transform shape] at (\offsetEone+90,-\bargap) {fbars={53.76,26.88}};
\pic[gray!50,transform shape] at (\offsetEone+103,-\bargap) {fbars={241.94,235.99}};
\pic[thin,gray!50,transform shape]
  {rcurves={\offsetEtwo}{108}{78.79}{9.09}{12.12}{\ratesEtwo}};
\pic[gray!50,transform shape] at (\offsetEtwo+105,-\bargap) {fbars={80.65,134.41}};
\pic[gray!50,transform shape] at (\offsetEtwo+118,-\bargap) {fbars={457,569.29}};
\end{tikzpicture}
\caption{In-game penalty frequencies and three-way outcome rates in dataset~D3.
For each minute of match time, the 5-minute rolling average of
{\color{blue}goal}, {\color{green}save}, and {\color{orange}miss}
rates is shown.
Since there is no information on stoppage time in our data,
we estimated the true frequency of penalties in the~45th and~90th minute
from those of the preceding 44~minutes
and spread penalties labeled with the last minute
at a constant rate into stoppage.
For better comparison, frequency bars in extra time are scaled 
by the ratio of matches with and without extra time~(1:26.88),
and rates clearly fluctuate because observations are too few.}
\label{fig:timeseries}
\end{figure}
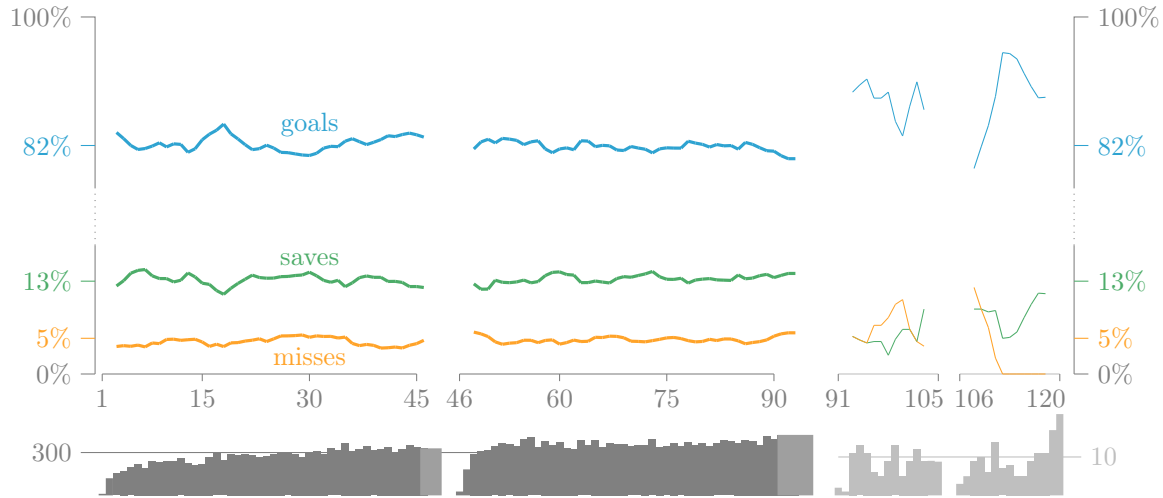

We first confirm previous observations 
that the rate at which penalties are awarded 
is increasing over match time \citep{horn:2021,dalton:2015}. 
Out of 30,908~regulation penalties in~D3,
13,044~(42.20\%) were awarded in the first half and
17,793~(57.57\%) in the second.%
\footnote{There are 71~in-game penalties for which time is not available.}
There is a noticably low number of penalties
in the first two minutes after kick-offs.
Since we have no information on the stoppage time in our data,
we extrapolated the number of penalties for an estimated stoppage time of three (five) minutes for the first (second) half. 

As already apparent from \Cref{fig:outcome_all},
data is very sparse in extra time,
and rates fluctuate strongly. 
This re-iterates how small data sets
can lead to wildly different results.
In regulation, on the other hand, rates are very stable over time
and even the difference between the first half~(82.41\%)
and the second half~(81.71\%) is small.

We conclude that empirical evidence does not support the idea
of fatigue having a noticeable effect on penalty conversion rates.

\subsection{First-mover advantage?}\label{sec:fma}

As fatigue is not plausible empirically,
the heightened stress accompanying these decisive actions
appears to be a major contributor to lower goal percentages
in the shootout condition. 

Stress affecting penalty takers is at the core of debates
around the order in which penalties are taken during shootouts.
Since penalties are more likely to be converted than not,
the team following in the alternating order is more often lagging behind.
Loss aversion and the prospect of not being able to level
would increase stress, and thus lead to an advantage for the team starting.
A number of studies observed this so-called first-mover advantage 
\citep{apesteguia:2010,palacios:2014,rudi:2020,dasilva:2018},
leading to the development of theoretical arguments and the proposal of alternatives
\citep{csato:2020,lambers:2020,vandebroek:2016,csato:2022,brams:2018}.
Other studies, however, rebuff the existence of a first-mover advantage
\citep{kocher:2012,kassis:2021,arrondel:2019,santos:2022}.

It should be noted that most of these studies are based on small datasets
stretched over very long periods of time.
The largest dataset to date covers~$1,623$~shootouts
in national and international competitions from~1970 to~2018~\citep{rudi:2020}.
According to this data,
which spans almost five decades,
the starting team wins~$54.71\%$ of all shootouts.

In contrast, our dataset~D2 contains $1,759$~shootouts
from the past eleven~seasons in competitions all across Europe.
According to our data, the starting team wins~$48.83\%$ of the shootouts.
Restricting attention to national cups associated with the top five leagues,
the teams going first win~$46.98\%$ of $513$~shootouts.
Hence, there is no empirical evidence for a first-mover advantage.
Either there is none, or teams have learned to compensate for it.

\begin{figure}
\centering
\edef\percentages{%
77.20/1759, 76.29/1759,
72.54/1759, 75.90/1759,
75.04/1759, 74.02/1759,
72.81/1736, 73.52/1624,
74.62/1379, 75.85/998,
74.53/530, 76.60/530,
78.19/321, 69.16/321,
73.00/200, 70.50/200,
61.90/105, 80.00/105}
\centerline{\begin{tikzpicture}[scale=0.07]
\draw[gray] (-2,0) node[left]{$0\%$} -- (0,0) -- (0,100) -- (-2,100) node[left]{$100\%$};
\draw[blue] (0,75) -- +(-2,0) node[left]{$75\%$};
\foreach \x in {10,12,...,18}
 {\draw[gray,dashed] (10*\x+5,0) -- +(0,100);}
\foreach \p/\n [count=\x] in \percentages
 {\pgfmathparse{isodd(\x)?"black":"red"}
  \colorlet{dotcolor}{\pgfmathresult}
  \fill[dotcolor] (10*\x,\p) circle (\n^0.5/20)
   +(60:1.5) node[above,rotate=60,anchor=west]{\footnotesize$\p\%$};
  \fill[gray] (10*\x-2.5,0) rectangle +(5,\n/100);
  \draw[gray] (10*\x,\n/100) node[above]{\scriptsize \n};
  \draw[dotcolor] (10*\x,40)
    node {\pgfmathparse{isodd(\x)?"A":"B"}\pgfmathresult
          \pgfmathparse{int((1+\x)/2))}\pgfmathresult};
 }
\end{tikzpicture}}
\caption{Goal rate by position in 1,759 shootouts from dataset~D2.
Dot areas correspond to the number of penalties given at the bottom.}
\label{shootout conversion}
\end{figure}
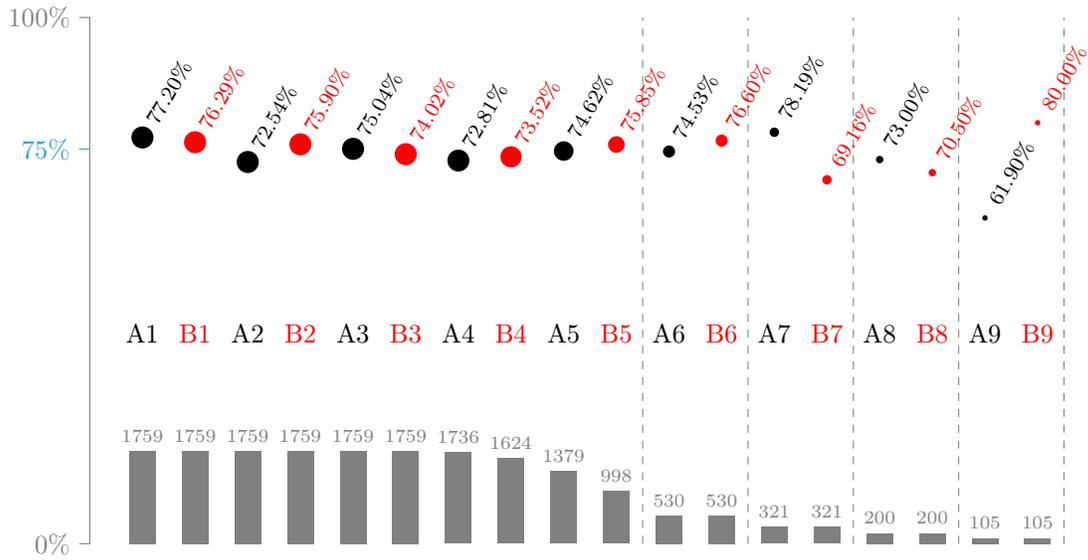

Similar to prior research
\citep{jordet:2007,arrondel:2019,dalton:2015,apesteguia:2010},
we next depict success rates for penalty kickers 
split by position in the shooting order
in \Cref{shootout conversion}.

Very few shootouts are decided after the minimum of six penalties,
the median number is ten, and a third require more than ten.
All but 56~shootouts in~D3 are decided after 18~or fewer penalties.
The maximum of 32~penalties was required in three shootouts:
Scunthorpe United beat Worcester City 14--13 in~2014 (English FA Cup),
Derby County beat Carlisle United 14--13 in~2016 (English Football League Cup), and
Stade Brestois~29 beat Dinan L\'ehon~FC 13--12 in~2021 (Coupe de France).

On average, the players taking the first two penalties have the highest goal rates.
It appears to be a common team strategy
to let the most reliable kickers go first~\citep{mcgarry:2000},
This is corroborated by the observation that
the in-game conversion rate of these players is also above average~($83.99\%$).
The second player of the starting team~(A2) has an exceptionally low rate,
and this is largely independent of the score:
($71.47\%$, $71.77\%$, or $72.69\%$ if the score is 0-1, 1-0, or 1-1,
and $76.77\%$ in 99~cases where it is 0-0).
The fifth kicker of the following team~(B5)
does not get to take a penalty in $43.26\%$ of all shootouts.
Since the number of shootouts halves with each subsequent pair of penalties,
we abstain from interpreting the increasingly fluctuating rates thereafter.

\begin{figure}
\edef\cumulative{
0.006239466/0.006695851/0.007959068/0.005116543,
0.052679211/0.024927595/0.044911882/0.018760659,
0.102839554/0.109331816/0.048891416/0.090392268,
0.273731212/0.169154339/0.167140421/0.049459920,
0.345989837/0.364765100/0.069357590/0.196702672,
0.345989837/0.364765100/0.000000000/0.000000000,
0.396424950/0.421203412/0.056281978/0.062535532,
0.396424950/0.421203412/0.000000000/0.000000000,
0.440405044/0.448707564/0.042637862/0.026151222,
0.440405044/0.448707564/0.000000000/0.000000000,
0.464284644/0.469814979/0.028425242/0.025582717,
0.464284644/0.469814979/0.000000000/0.000000000,
0.472443738/0.489898904/0.008527572/0.019329164
}
\begin{tikzpicture}[xscale=0.75,yscale=3]
\foreach \i in {1,...,9} 
 {
  \draw[black] (2*\i-0.5,-0.6) node{A\i}; 
  \draw[red] (2*\i+0.5,-0.6) node{B\i}; 
 }
\path (6,0.5) coordinate (A) coordinate (Ba)
      (6,-0.5) coordinate (B) coordinate (Ab);
\foreach \pa/\pb/\fa/\fb [count=\n] in \cumulative
 {
  \fill[black!30] (5+\n, 0.5) rectangle +(1,-\pa);
  \fill[red!30]   (5+\n,-0.5) rectangle +(1, \pb);
  \draw[black] (A) -- ++(0,-\fa) -- ++(1,0) coordinate (A);
  \draw[red]   (B) -- ++(0, \fb) -- ++(1,0) coordinate (B);
  \draw[red,dotted]   (Ba) -- ++(0,-\fb) -- ++(1,0) coordinate (Ba);
  \draw[black,dotted] (Ab) -- ++(0, \fa) -- ++(1,0) coordinate (Ab);
 }
\draw[gray,thick,dashed]
  foreach \n in {11,13,15,17} {(\n,-0.5) -- +(0,1)};
\node[black,right] at (15.25, 0.25) {A win};
\node[red,right]   at (15.25,-0.25) {B win};
\draw[gray,thin] (1,0) -- +(18,0) node[right]{50:50};
\draw[gray] (1,-0.5) rectangle +(18,1);
\end{tikzpicture}
\caption{Winning rates within first 18~penatlies
of 1,759~shootouts in dataset~D2.
Background bars are cumulative winning probabilities derived
from independent conversion probabilities 
according to rates in \Cref{shootout conversion}.
Solid lines of actual case numbers match them almost perfectly,
and are mirrored as dotted lines for better comparison.}
\label{shootout winning}
\end{figure}
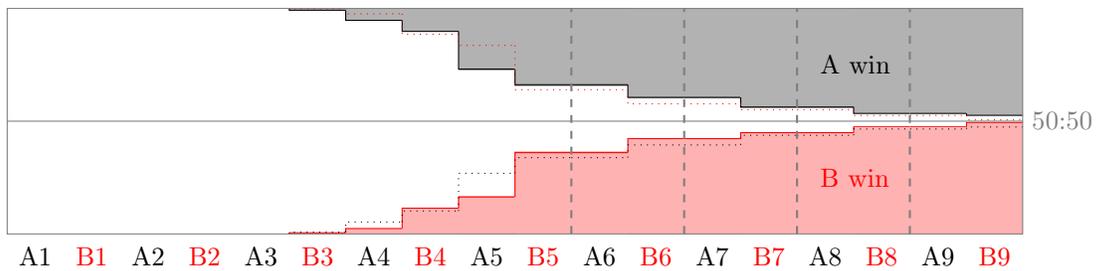

The cumulative winning rates of both teams
are depicted by solid lines in \Cref{shootout winning}.
For comparison, we also calculated the cumulative rates at which teams win,
if penalties are converted independent from each other
with the observed relative frequencies
for each position in the shooting order
shown in \Cref{shootout conversion}.
The close match of these two curves suggests
that scoreline effects, if any, cancel each other out.
The starting team (Team~A) tends to win earlier, if they win,
but as noted above they do not win more often overall.
Teams win most often by converting their fifth penalty:
Of all shooutouts in~D2,
17\% are won by Team~A on converting~A5
and 20\% are won by Team~B on converting~B5.

In fact, as summarized in \Cref{last table},
two thirds of all shootouts are decided on a goal
(because the other team cannot draw level with the remaining attempts).
Note that this is conditional on the fact that this penalty did end the shootout,
and therefore different from the conversion rate of a potentially last penalty.
The intuitive argument of increased pressure affecting kicker performance,
is not supported by our data, as there is virtually no difference in performance 
between potentially deciding penalties and all others.

\begin{table}
\caption{Factually and potentially last penalties compared to others in~D2.}
\begin{tabular}{lrr}
 situation & penalties & goal \\ \hline
 last                &  1,759 & $65.55\%$ \\ 
 not last            & 17,144 & $75.57\%$ \\ \hline
 potentially winning &  1,522 & $75.76\%$ \\ 
 potentially losing  &  2,371 & $74.44\%$ \\ 
 neither             & 15,010 & $74.56\%$ \\ 
\end{tabular}
\label{last table}
\end{table}  

\begin{table}
\caption{Relative frequencies of outcomes
for in-game penalties in~D3 split by current score. 
(Information on the current score is missing for 15~penalties.)}
\begin{tabular}[b]{lrrrr}
score & penalties & goal      & save      & miss  \\ \hline
in front &  8,567 & $82.37\%$ & $12.85\%$ & $4.77\%$\\ 
level    & 13,294 & $82.02\%$ & $13.19\%$ & $4.78\%$\\ 
behind   &  9,337 & $81.58\%$ & $13.69\%$ & $4.73\%$\\ 
\end{tabular}
\label{score outcome}
\end{table} 

In a final comparison, we look at success rates of in-game penalties 
in relation to the current score (\Cref{score outcome})
and find no significant difference either.

It appears, therefore, that the shootout condition as a whole, 
with its decisive character, heightened attention, and long walk to the spot,
is putting pressure on penalty takers, not the current score.

\section{Conclusion}\label{sec:conclusion} 

We descriptively analysed more than 50,000~penalties
from top-division leagues and cups
across the Union of European Football Associations~(UEFA)
over a period of eleven~seasons from 2012/2013 to 2022/2023.
Approximately one third of these are shootout penalties.

Since raw data of more than 60,000~penalties
has been obtained from the unofficial records of Transfermarkt,
statistical tests based on suspected types of inconsistencies
were employed to increase data quality
before detailed analyses were performed.
For readability and simplicity, we summarize our findings as follows:
\begin{itemize}
\item The rate at which penalties are awarded is increasing over the course of a match.
\item Out of 20~in-game penalties, 16~are converted, three saved and one missed. 
\item Out of 20~shootout penalties, 15~are converted, three saved and two missed.
\item The goal rate of regular penalty takers
is four percentage points lower in shootouts.
This drop in performance is accounted for by additional misses 
and may therefore be due to pressure on the kicker 
rather than to heroics of the goalkeeper.
\item The goal rate of non-regular penalty takers appearing only in shootouts
is lower by a another four and a half percentage points.
This additional drop is accounted for equally by saves and misses
and may therefore be due to player ability.
\item In shootouts, conversion rates suggest 
that both teams let their most reliable kickers go first,
but also that starting teams select their second kicker differently.
\item More than four out of ten shootouts are decided
before the fifth kicker of the second team gets to take his penalty.
\item There is no first-mover advantage in shootouts, or it is compensated for.
Either way, the alternating shooting order seems fair.
\item Two thirds of penalty shootouts are decided by
a team succeeding rather than failing on the final penalty.
\end{itemize}

We acknowledge several limitations to our study.
Despite our best efforts to account for inconsistencies and reporting bias, 
we may have been to eager to remove seeming outliers 
and there may be other types of errors that we miss. 
Both could lead to systematic biases in the outcome rates we report.
Since the overall results are consistent with those in top leagues
(where data quality can be expected to be better because of higher attention) 
we do not expect this problem to be a threat to validity.

We have repeatedly emphasized that we do not attempt to
analyze techniques and strategies to increase penalty success.
Further limitations in scope include our focus on the past eleven~seasons.
While the data is more recent than in other studies, 
we cannot make any statement about historical developments
or the affect of recent rule changes. 
In top-level competitions,
the now extensive scrutiny of goalkeepers staying on the line
might work against them and lead to increased goal rates.
Except for the FIFA World Cup Finals, 
we also do not cover competitions outside of UEFA,
and we have deliberately excluded women's football. 
Extensions in either direction may lead to qualitatively different insights.

In should be borne in mind that we have described large-scale empirical phenomena,
and it is at the heart of the attractiveness of the game
that anything can happen in single matches or shootouts,
and with single players or teams.
We hope that our study can serve as a baseline
to set expectations in modern-day association football. 
As such it could be helpful for practioners, commentators, and viewers alike.


\end{document}